\documentclass[useAMS,usenatbib]{mn2e}
\usepackage{aasmacros}
\usepackage{times}
\usepackage{graphicx}
\usepackage{amssymb}
\usepackage[usenames,dvipsnames]{color}
\usepackage{url}
\usepackage{fixltx2e}
\usepackage{mathtools}
\usepackage{amsmath}
\usepackage{float}
\usepackage{cite}
\usepackage{fixltx2e}
\bibpunct{(}{)}{;}{a}{,}{,}
\usepackage{longtable}
\usepackage{hyperref}
\usepackage{morefloats}

\let\cite=\citen

\def\apjs{ApJl}

\def\pd{\textit{P(D)}}

\defcitealias{Vernstrom13}{V14}
\bibpunct{(}{)}{;}{a}{}{,}

\title[The Deep Diffuse Extragalactic Radio Sky at 1.75 GHz]{The Deep Diffuse Extragalactic Radio Sky at 1.75 GHz}

\author[Vernstrom et. al]{T. Vernstrom$^1$\thanks{E-mail:tvern@phas.ubc.ca}, Ray P. Norris$^2$, Douglas Scott$^1$, J.V. Wall$^1$\\
  $^1$Department of Physics and Astronomy, University of British Columbia, Vancouver, BC V6T 1Z1, Canada\\
  $^2$Australia Telescope National Facility, PO Box 76, Epping NSW 1710, Australia\\
}
\begin{document}
  

\pagerange{\pageref{firstpage}--\pageref{lastpage}} \pubyear{2014}

\maketitle

\label{firstpage}
\begin{abstract}
We present a study of diffuse extragalactic radio emission at $1.75\,$GHz from part of the ELAIS-S1 field using the Australia Telescope Compact Array. The resulting mosaic is $2.46\,$deg$^2$, with a roughly constant noise region of $0.61\,$deg$^2$ used for analysis. The image has a beam size of $150 \times60\,$arcsec and instrumental $\langle\sigma_{\rm n}\rangle= (52\pm5)\, \mu$Jy beam$^{-1}$. Using point-source models from the ATLAS survey, we subtract the discrete emission in this field for $S \ge 150\, \mu$Jy beam$^{-1}$. Comparison of the source-subtracted probability distribution, or \pd, with the predicted distribution from unsubtracted discrete emission and noise, yields an excess of $(76 \pm 23) \, \mu$Jy beam$^{-1}$. Taking this as an upper limit on any extended emission we constrain several models of extended source counts, assuming $\Omega_{\rm source} \le 2\,$arcmin. The best-fitting models yield temperatures of the radio background from extended emission of $T_{\rm b}=(10\pm7) \,$mK, giving an upper limit on the total temperature at $1.75\,$GHz of $(73\pm10)\,$mK. Further modelling shows that our data are inconsistent with the reported excess temperature of ARCADE2 to a source-count limit of $1\, \mu$Jy. Our new data close a loop-hole in the previous constraints, because of the possibility of extended emission being resolved out at higher resolution. Additionally, we look at a model of cluster halo emission and two WIMP dark matter annihilation source-count models, and discuss general constraints on any predicted counts from such sources. Finally, we report the derived integral count at $1.4\,$GHz using the deepest discrete count plus our new extended-emission limits, providing numbers that can be used for planning future ultra-deep surveys.

\end{abstract}


\section{Introduction}
\label{sec:introduction}

It is well known that the bulk of discrete radio sources, for frequencies near $1.4\,$GHz, is made up of either active galactic nuclei (AGN) or starburst galaxies. Their space distributions have been modelled via their luminosity functions to determine how they evolve with redshift. The number counts ${\rm d}N/{\rm d}S$, or the differential number of sources per steradian per flux density interval, have been measured well into the sub-mJy region \citep{Owen08,Condon12}, as well as the contribution from these sources to the cosmic radio background (CRB). However, what is less well characterized is the extended large-angular-scale emission associated with these galaxies or perhaps associated with groups or clusters of galaxies or with the cosmic web. There have been few surveys carried out for diffuse arcmin-scale extragalactic emission, and very few that also have high sensitivity at that scale. The most sensitive lower resolution survey yet published was by \citet{Subrah10}, which reached an rms of $85\, \mu$Jy beam$^{-1}$ with a $50\,$arcsec beam. This angular scale encompasses extended low-surface-brightness emission from individual galaxy haloes, as well as emission from the intra- and inter-cluster medium; features such as giant and mini radio haloes, radio relics, and diffuse emission from filaments and other structures in the inter-cluster medium \citep[see e.g.][]{Feretti12}. 

It is unknown how much this large-scale emission may contribute to the cosmic radio background (CRB) temperature. This background at radio frequencies ($T_{\rm b}$) is composed of emission from the cosmic microwave background (CMB, $T_{\rm CMB}$), the contribution from the Milky Way ($T_{\rm Gal}$), and the contribution from extragalactic sources ($T_{\rm source}$); thus $T_{\rm b}=T_{\rm CMB}+T_{\rm Gal}+T_{\rm source}$. The CMB contribution has been measured to high accuracy and corresponds to a blackbody with $T=2.7255\,$K \citep{Fixsen09a}. Recent estimates from the deep survey by \citet{Condon12} and \citet[hereafter V14]{Vernstrom13} were made of the contribution from extragalactic sources using the Karl G. Jansky Very Large Array (VLA) at $3\,$GHz. They found the contribution from compact sources to be $T_{\rm source}=14\,$mK at $3\,$GHz and $120\,$mK when scaling this result to $1.4\,$GHz. However, the beam size from the VLA at $3\,$GHz was $8\,$arcsec and the image was constructed from \textit{uv} weighting that filtered out scales much larger than the beam. Thus that survey would not have been sensitive to emission on larger scales.  

Extended low-surface-brightness radio emission can be difficult to survey. Galactic- and cluster-scale emission can extend up to several arcminutes. Single-dish telescopes at radio frequencies have beams on much larger scales and are limited in their continuum sensitivity by systematic errors, while most interferometers are not ideal for measuring low-surface-brightness extended objects. The surface brightness sensitivity of an interferometer is limited by its spatial frequency coverage in the image domain, which is the Fourier transform of its coverage of the aperture plane, often referred to as its `\textit{uv } coverage'. For example, if an interferometer consists of antennas of diameter $D$, and the length of the shortest baseline is $b$, then the interferometer is generally insensitive to objects in the sky with angular size greater than $\lambda/(b-D)$ radians. An interferometer with $D=25\,$m and $b=1000\,$m observing at $20\,$cm is therefore insensitive to astronomical objects with scale sizes greater than $0.7\,$arcmin. Mosaicing can recover spatial information $>{\lambda/D}$ in size but nothing can recover information between $>{\lambda/ D}$ and $<\lambda/(b-D)$, as that has not been measured by the interferometer. Thus, not many deep extended emission surveys have been carried out at radio frequencies.

The issue of large-scale emission and the CRB has been of greater interest in the last few years, following the results of ARCADE 2 \citep{Fixsen09,Fixsen11}. This balloon-borne experiment observed the sky at several radio frequencies, ranging from $3.3$ to $100\,$GHz. It measured a background temperature at $3.3\,$GHz that is much higher than current estimates from extragalactic sources, $(54\pm6)\,$mK compared with the $T_{\rm source}\simeq14\,$mK of \citetalias{Vernstrom13}. \citet{Singal10} proposed that the excess could be due to a new population of faint distant star-forming galaxies. \citetalias{Vernstrom13} ruled out any new populations of discrete compact sources having peaks in the source count above $50\,$nJy.

For compact sources to be causing the excess emission seen by ARCADE 2, the additional sources would need to have number-count peaks at very faint flux densities. This could raise a problem with the far-IR to radio correlation (unless this correlation evolves with redshift), and conflict with limits on the overall number of galaxies. 

However, the cause of the ARCADE 2 excess \textit{could} be larger-scale emission (scales ranging from around $0.5\,$arcmin up to the $12^{\circ}$ primary beam size of the ARCADE 2 experiment). It has been proposed that the emission could be caused by dark matter annihilation \citep{Fornengo11,Hooper12,Yang13,Fornengo14}, in which case it would trace the dark matter distribution of clusters of galaxies, with a characteristic scale size of arcmin. Other emission processes could include those normally seen from clusters, such as radio relics and haloes, or with diffuse synchrotron emission from the cosmic web \citep{BrownSD11}. Such emission processes do not directly correlate with star formation and therefore could evade constraints from the far-IR radio correlation.

In this paper we use deep low-resolution radio observations from the Australia Telescope Compact Array (ATCA) to investigate the emission that might be present at larger angular scales and constrain how it might contribute to the CRB. In Section~\ref{sec:obs} we detail the radio observations, data reduction, and imaging process, as well as discussing the image noise properties. Section~\ref{sec:pofd} describes the technique used to examine the data. In Section~\ref{sec:discrete} we discuss our treatment of discrete point sources, our subtraction method, and the contribution from faint un-subtracted sources. We discuss issues of detecting extended emission at both high and low resolutions in Section~\ref{sec:extcnt}. Section~\ref{sec:mods} details the models we use for investigating the extended or diffuse emission. Section~\ref{sec:btemp} discusses the conversion from source count to background temperature, as well as the predicted background temperatures from ARCADE 2 at our image frequency. Section~\ref{sec:results} presents the results of fitting our extended emission source count models to our new data and their contribution to the CRB, and discusses models fit to the ARCADE 2 results. In Section~\ref{sec:dis} we discuss our findings, in particular what the results might mean in terms of astrophysical sources. We examine models of cluster halo emission as well as a source count models from dark matter. Finally, in Section~\ref{sec:intc} we present our current estimates of integral source counts for both discrete and extended source count models.
 
\section{Radio Observations}
\label{sec:obs}

We targeted a portion of the European Large Area \textit{Infrared Space Observatory} Survey -- South 1 field \citep[ELAIS-S1,][]{Oliver00}, an extragalactic region originally selected for \textit{ISO} observations. This field was chosen because it has previously been surveyed with higher resolution for the Australia Telescope Large Area Survey \citep[ATLAS,][Franzen et. al, 2014 in preparation, Banfield et. al, 2014 in preparation]{Norris06, Middelberg08,Hales14a}. Our new observations were made with the ATCA EW352 array configuration, which has a maximum baseline of $352\,$m and a minimum baseline of $30.6\,$m. A total of 12 hours of observation time was obtained in a single session on November, 28 2013. We observed using the ATCA $13\,$cm band, which centres on $2.1\,$GHz with  $2\,$GHz of bandwidth. The bandwidth is separated into 2048 channels of $1\,$MHz width. The resolution with this configuration ranges from $1$ to $2\,$arcmin, depending on the image frequency. We observed seven pointings in the ELAIS-S1 field, chosen from the 20 pointings used by the ATLAS survey. The pointing centres are listed in Table~\ref{tab:point} and are shown in Fig.~\ref{fig:images}.
 
 \begin{table}
\caption{ATLAS ELAIS-S1 pointings.}
\begin{tabular}{lllll}
\hline
\hline
ATLAS & RA J2000 & Dec J2000 & \multicolumn{2}{c}{$\sigma_{\rm n}$}\\
Pointing & \scriptsize{(HH:MM:SS.ss)} & \scriptsize{(HH:MM:SS.ss)} & (${\rm \mu}$Jy beam$^{-1}$)& (mK)\\
 \hline
 el1$\_$1 & $00$:$32$:$03.6$&$-43$:$44$:$51.2$& $53.7 \pm 4.64$&$2.4\pm0.21$\\
 el1$\_$5 & $00$:$32$:$57.7 $&$-43$:$28$:$09.0$& $52.3 \pm 2.66$&$2.3\pm0.12$\\
 el1$\_$6 & $00$:$33$:$50.8 $&$-43$:$44$:$57.4$& $57.0 \pm 5.31$&$2.5\pm0.24$\\
 el1$\_$7 & $00$:$35$:$38.0 $&$-43$:$44$:$57.4$& $57.8 \pm 3.18$&$2.6\pm0.14$\\
 el1$\_$8 & $00$:$34$:$44.4 $&$-43$:$28$:$11.9$& $58.1 \pm 7.28$&$2.6\pm0.32$\\
 el1$\_$16 &$00$:$34$:$44.4$&$-44$:$01$:$42.8$& $59.2 \pm 6.51$&$2.6\pm0.29$\\
 el1$\_$17 &$00$:$32$:$57.7$&$-44$:$01$:$42.8$& $50.8 \pm 3.61$&$2.3\pm0.16$\\
 \hline
 \end{tabular}
 \label{tab:point}
 \end{table}
 
\subsection{Calibration and editing}
\label{sec:edit}
The calibration, editing, and imaging were performed using the \textsc{miriad}\footnote{\url{http://www.atnf.csiro.au/computing/software/miriad/}} software package \citep{Salt95}. Following several rounds of RFI flagging, the source J1934$+$638 was used for bandpass and flux density calibration. The source PKS 0022$-$423 (PMN J0024$-$4202) was observed for $2\,$-minute intervals every $10\,$minutes and used to correct the gain phases. The task \textsc{gpcal} was utilized to derive frequency-dependent gain solutions, solving for the gains of the upper and lower parts of the band separately. 

Observations at this frequency are highly affected by radio frequency interference (RFI), most notably at the lowest frequencies. The task \textsc{mirflag} was used for automated RFI flagging on the phase calibrator source and the target fields. This allowed us to flag the majority of interference, so that only a small amount of manual flagging was required. Each of the seven pointings was flagged individually for {\it uv} points above an amplitude threshold. The data were then split into two frequency bands (1.1 to $2.1\,$GHz and 2.1 to $3.1\,$GHz), and separated into individual data sets for each pointing. The last hour of time was not usable, since the source was setting, and for the the final four hours Antenna~1 was lost due to shadowing. In the end about $55\,$per cent of the data was flagged (i.e. not used) in the $1.1$ to $2.1\,$GHz frequency band, and about $30\,$per cent in the $2.1$ to $3.1\,$GHz band.

The following analysis is only carried out for the lower part of the band  ($1.1$ to $2.1\,$GHz), which, after flagging, ranged from $1.38$ to $2.1\,$GHz, with a centre frequency of $1.75\,$GHz. This decision was made because it more closely matches the image frequency of the ATLAS survey. We planned to use the ATLAS point-source models to subtract discrete emission from our data. The spectral change of the primary beam going from the lower band to the upper band ($2.1$ to $3.1\,$GHz) is large, which makes accurate scaling of the point source models difficult and the output of the subtraction at the higher frequency less reliable. For this reason we do not believe the addition of the upper band would contribute additional meaningful information for our analysis.

\subsection{Imaging}
\label{sec:imaging}
Imaging was first performed on the full \textit{uv} data sets, primarily for the purposes of self-calibration of the data. However, the ultimate goal was to perform subtraction of the known point sources in the fields and re-image the source-subtracted data for further analysis. The subtraction process is discussed in more detail in Section~\ref{sec:subtract}.

The \textsc{miriad} tasks \textsc {invert}, \textsc{mfclean}, and \textsc{restor} were used to create and clean the images. Due to the large frequency range covered, we used multi-frequency synthesis and deconvolution, or cleaning (\textsc{mfclean}). \textsc{mfclean} attempts to solve for a frequency dependent intensity, $I(\nu)$. Here
\begin{equation}
I(\nu)=I(\nu_{0})\left(\frac{\nu}{\nu_{0}}\right )^{\alpha},
\label{eq:inu}
\end{equation}
and solving for the partial derivative of the intensity with frequency gives the spectral index,
\begin{equation}
I(\nu_{0})\alpha=\nu_{0}\frac{\partial I}{\partial \nu} \Big \arrowvert_{\nu_{0}}.
\label{eq:inual}
\end{equation}
Thus by using \textsc{mfclean} the resulting image has two planes, the intensity at the reference frequency and the intensity times the spectral index. This allows us to take advantage of the large bandwidth and solve for the frequency dependence of sources (though a high signal-to-noise ratio is usually required in order to produce an accurate measurement). Note that this process can be complicated by the changing primary beam size at the different frequencies. There should therefore be an additional term representing the spectral dependence of the primary beam; however, currently \textsc{mfclean} only allows for fitting of one additional spectral term. Instead the primary beam frequency dependence was accounted for during the mosaicing process.

\begin{figure*}
\includegraphics[scale=0.53,natwidth=12in,natheight=12in]{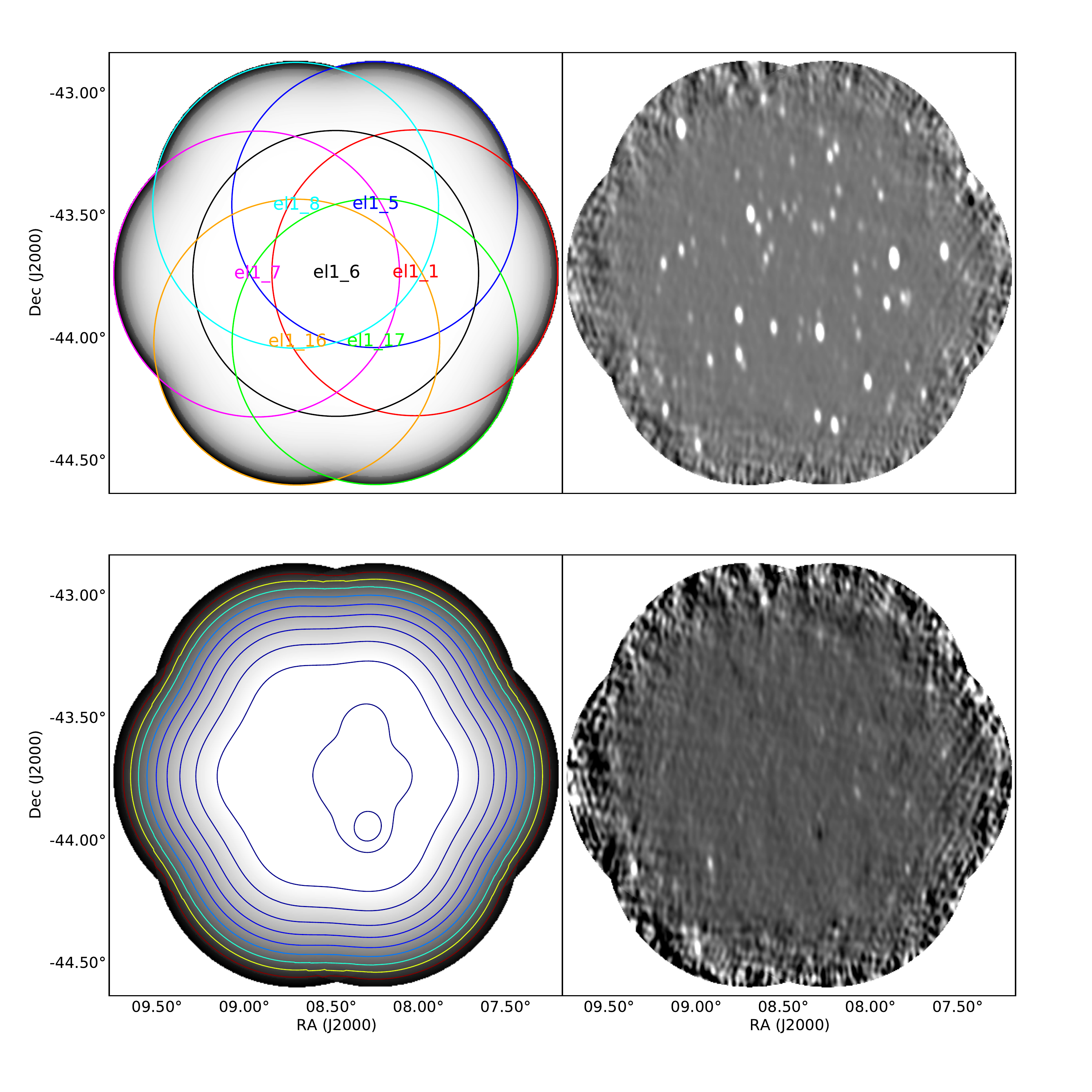}
\caption{ELAIS-S1 mosaics. The top left panel shows the full area, with the seven pointings outlined and labelled at their centres. The top right panel is the final $1.75\,$GHz mosaic image. The bottom left panel shows the noise across the mosaic field, with contour levels at $46, 48, 78,120, 190, 305, 480, 760, 1200, 1900,$ and $3000 \,{\rm  \mu}$Jy beam$^{-1}$. The bottom right panel shows the image after subtraction of the ATLAS point sources. }
\label{fig:images}
\end{figure*}

Each pointing was cleaned separately, initially down to a level of $600\, \mu$Jy beam$^{-1}$. At this stage we performed two rounds of phase-only self-calibration and one of amplitude and phase. The final images were cleaned down to $250\, \mu$Jy beam$^{-1}$. The resulting synthesized clean beam\footnote{In radio interferometry images the point spread function (PSF) resulting from the Fourier transform of the \textit{uv} coverage and weighting functions is known as the `dirty' synthesized beam. The dirty beam generally contains positive and negative sidelobes. The cleaning process finds bright peaks and stores them in a model as pixel flux densities known as the `clean' model. These model components are then convolved with the central Gaussian of the dirty beam, which is known as the clean synthesized beam, and added back to the original image. This clean beam is free of sidelobes.} size is $150\,$arcsec$\times60\,$arcsec, with a position angle of $6^{\circ}$, using Briggs\footnote{Uniform weighting of the \textit{uv} data usually results in a better behaved synthesized beam, and smaller side lobes, but usually with higher noise. Natural weighting generally gives the best signal-to-noise ratio (though not in the confusion-limited case), but at the expense of an increased beam size. Briggs, or `robust', weighting allows for weighting between the two options, doing so in an optimal sense (similar to Wiener optimization).} weighting \citep{Briggs95}  and a robustness factor of $0.5$. A mosaic of the seven pointings was made using \textsc{linmos}, with each pointing having a primary beam full width at half maximum (FWHM) of roughly $27\,$arcmin; the final mosaic (Fig.~\ref{fig:images}) has a total area of approximately $2.46\,$deg$^2$. 

Regarding the brightness units, radio images are often generated with brightness units of Jy beam$^{-1}$. However, it can be useful (particularly with this type of discussion) to convert these units to that of brightness temperature in K (or mK). Conversion between these units can be computed using a factor 
\begin{equation}
\mathcal{C}_{\rm T}=\frac{\lambda^2 10^{-26} {\rm Wm}^{-2}}{2 k_{\rm B}\Omega_{\rm beam}},
\label{eq:tconv}
\end{equation}
such that $T=\mathcal{C}_{\rm T}S$ in Kelvin with $S$ the flux density in units of Jy beam$^{-1}$, and where $\Omega_{\rm beam}$ is the beam solid angle in steradians, and $k_{\rm B}$ is the Boltzmann constant. Thus for our case of a Gaussian ellitpical beam with FWHM sizes of $150\,$arcsec$\times60\,$arcsec and frequency of $1.75\,$GHz, $\mathcal{C}_{\rm T}=44.54$. Throughout the rest of this paper, for convenience, we present results in both units.

\subsection{Image noise}
\label{sec:noise}

\begin{figure*}
\includegraphics[scale=0.5,natwidth=14in,natheight=5.6in]{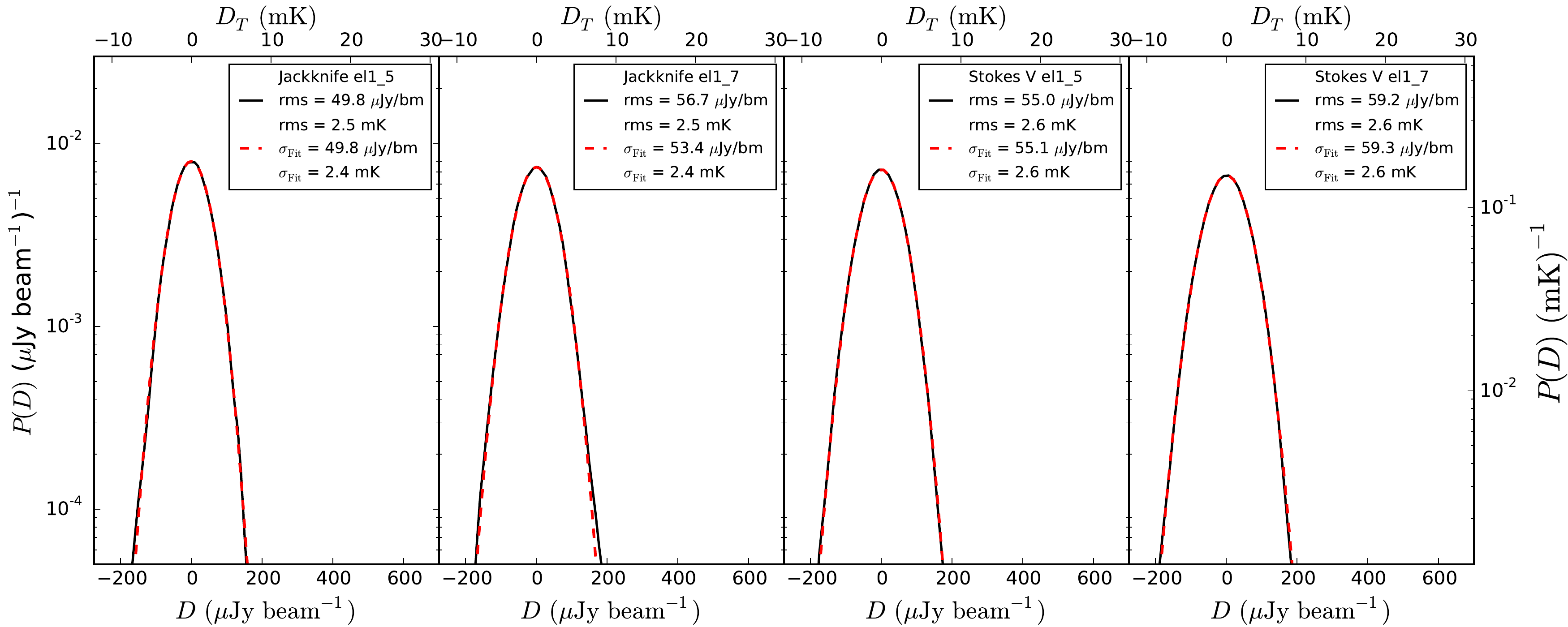}
\caption{Measurements and estimates of the instrumental noise using the jackknife method (first two panels) and Stokes $V$ method (second two panels) for two of the pointings. The black solid lines are from the pixel histograms of the images and the red dashed lines are fitted Gaussians. The quoted values are the measured rms from the image pixel values, while $\sigma_{\rm Fit}$ is the width of the fitted Gaussian.}
\label{fig:jackn}
\end{figure*}

Obtaining a precise measure of the instrumental noise $\sigma_{\rm n}$ is difficult, because with the large beam size the confusion rms $\sigma_{\rm c}$ is expected to dominate over the instrumental noise. However, for our analysis goals an accurate measurement and characterization of the noise is required. We employed two different techniques in order to estimate the instrumental image noise.  First we made measurements of the noise using the ``jackknife'' method. This involves taking two (approximately) equal halves of the data and creating separate images. Each of these images should have noise equal $\sqrt{2}\sigma_{\rm total}$. By differencing the images and dividing by two, the noise of the combined image should be left, with all the  signal subtracted out. Since the noise in each half adds in quadrature, then after the subtraction,
\begin{equation}
\sigma=\frac{\sqrt{\sigma_1^2+\sigma_2^2}}{2}=\frac{\sqrt{(\sqrt{2}\sigma_{\rm total})^2+(\sqrt{2}\sigma_{\rm total})^2}}{2}=\sigma_{\rm total}.
\label{eq:jackk}
\end{equation}

It can be challenging with interferometry to create images with equal halves of the data. Choosing two equal time chunks can introduce issues with different \textit{uv} coverage between the two data sets. We therefore chose to create two images using the even and odd numbered spectral channels, which should give approximately half in each set. The images were cleaned in the same manner and then subtracted for each pointing. We measured the rms in the cleaned portion of the image, as well as fitting the pixel distribution with a Gaussian to obtain a fitted rms noise $\sigma_{\rm n}$. This can be seen for two of the pointings in Fig.~\ref{fig:jackn}. The jackknife procedure yielded measurements of the instrumental noise of the individual pointings of $50$--$65\, {\rm \mu}$Jy beam$^{-1}$, or $\,2.2$--$2.9\,$mK.

We used a second approach as a check on this procedure. The Stokes $V$ parameter measures circular polarization and is defined as
\begin{equation}
V=\langle E_{\rm l}^2 \rangle - \langle E_{\rm r}^2 \rangle,
\label{eq:stokesv}
\end{equation}
where $E_{\rm l}$ and $E_{\rm r}$ are, respectively, the left and right hand complex electric field amplitudes in the circular basis as measured by the antennas. The total intensity, or the Stokes $I$ parameter, is defined as 

\begin{equation}
I=\langle E_{\rm l}^2 \rangle + \langle E_{\rm r}^2 \rangle.
\label{eq:stokesi}
\end{equation}
Extragalactic radio sources generally have low levels of circular polarisation \citep{Rayner00} and so a Stokes $V$ image should have subtracted out all the signal, leaving only instrumental noise (similar to the jackknife, but performed in the \textit{uv} plane rather than the image plane). We therefore made Stokes $V$ images of all the pointings and again measured the rms and fit Gaussians to the pixel probability distributions to obtain a fitted rms $\sigma_{\rm n}$. This yielded similar estimates of $55$--$65\, {\rm \mu}$Jy beam$^{-1}$ ($\,2.4$--$2.9\,$mK), as can also be seen in Fig.~\ref{fig:jackn}. For final values of $\sigma_{\rm n}$ we averaged the measured and fitted values from the jackknife and Stokes $V$ for each pointing, and have listed them in Table~\ref{tab:point}. These values only account for instrument noise and do not include any additional noise contributions from the imaging process, such as uncleaned dirty beam sidelobes, artefacts from bright sources, or from sources out in the lobes of the primary beam (of which there are several).

\begin{figure}
\includegraphics[scale=0.475,natwidth=7in,natheight=7in]{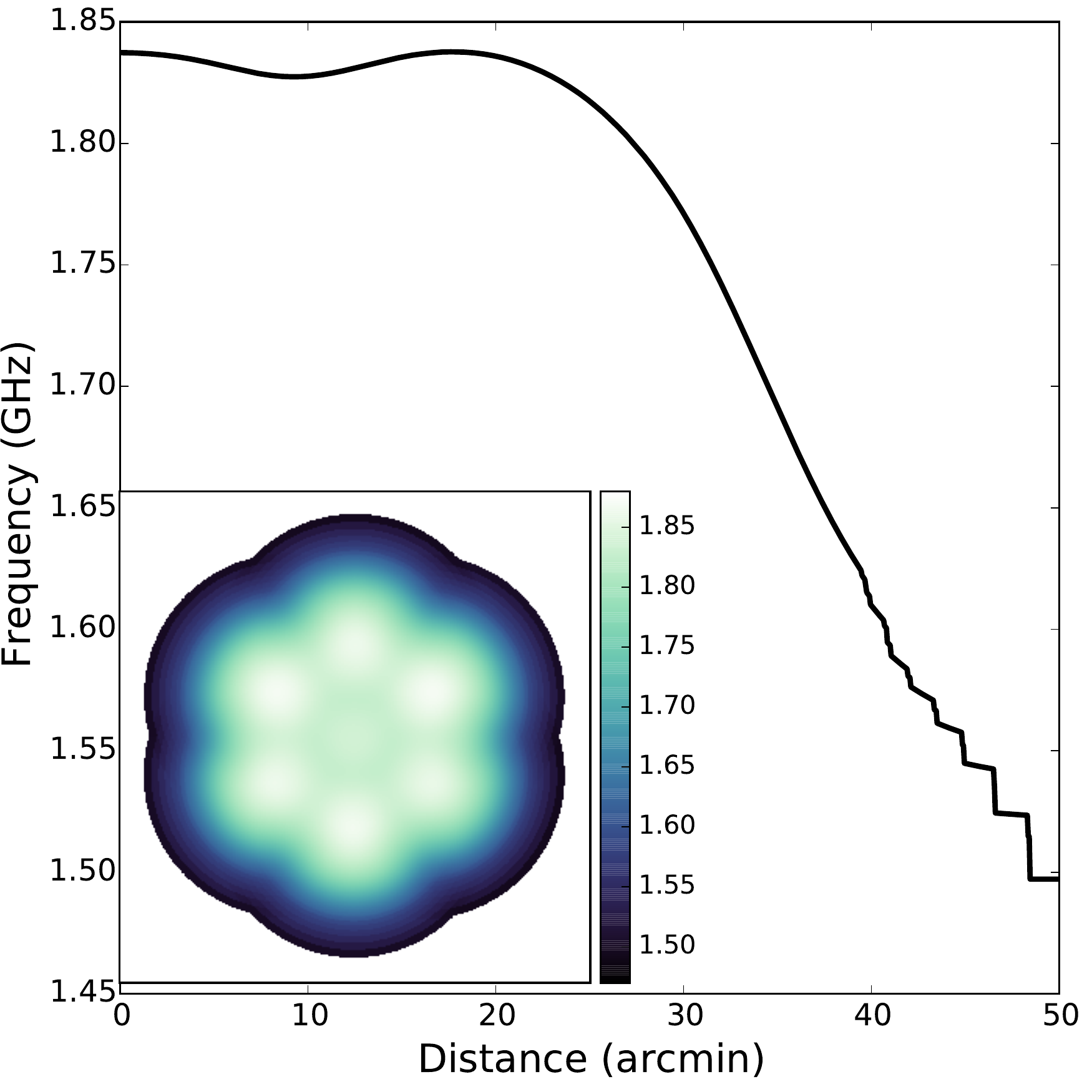}
\caption{Frequency dependence of the final mosaic image due to the wide-band primary beam correction.  The solid black line shows the effective frequency $\langle \nu \rangle$ from the centre to the edge of the image, as a function of radius. The inset is the full mosaic image, with the colour scale showing the change in effective frequency.}
\label{fig:freq}
\end{figure}

For the final mosaic, each pointing had a primary beam correction applied to it, raising the noise radially. \textsc{linmos} takes in the values of $\sigma_{\rm n}$ for each pointing and combines pixels by weighting as
\begin{equation}
S(x,y)=\sum_i \frac{S_i(x,y)}{(\sigma_{{\rm n},i}/ p_i(x,y))^2},
\label{eq:linmos}
\end{equation} 
where $S(x,y)$ is the final flux density of the pixel, $S_i(x,y)$ is the flux density in pointing $i$, $\sigma_{{\rm n},i}$ is the noise in pointing $i$ and $p_i(x,y)$ is the primary beam correction of pointing $i$ at position $(x,y)$. This results in non-uniform noise across the field. The resulting instrumental noise for the full mosaic is shown with contours in Fig.~\ref{fig:images}. The actual procedure used to combine the pointings is more complicated than eq.~(\ref{eq:linmos}), since, due to the wide bandwidth, the primary beam correction becomes frequency dependent. \textsc{linmos} takes into account the bandwidth used, as well as the spectral index information found from \textsc{mfclean}, to correct for the frequency effects. This results in an effective frequency $\langle \nu \rangle$ in the field that varies with distance from the centre, going from $1.75 $ to $1.4\,$GHz, as shown in Fig.~\ref{fig:freq}.

\section{Probability of Deflection}
\label{sec:pofd}
\begin{figure*}
\includegraphics[scale=.55,natwidth=13in,natheight=4in]{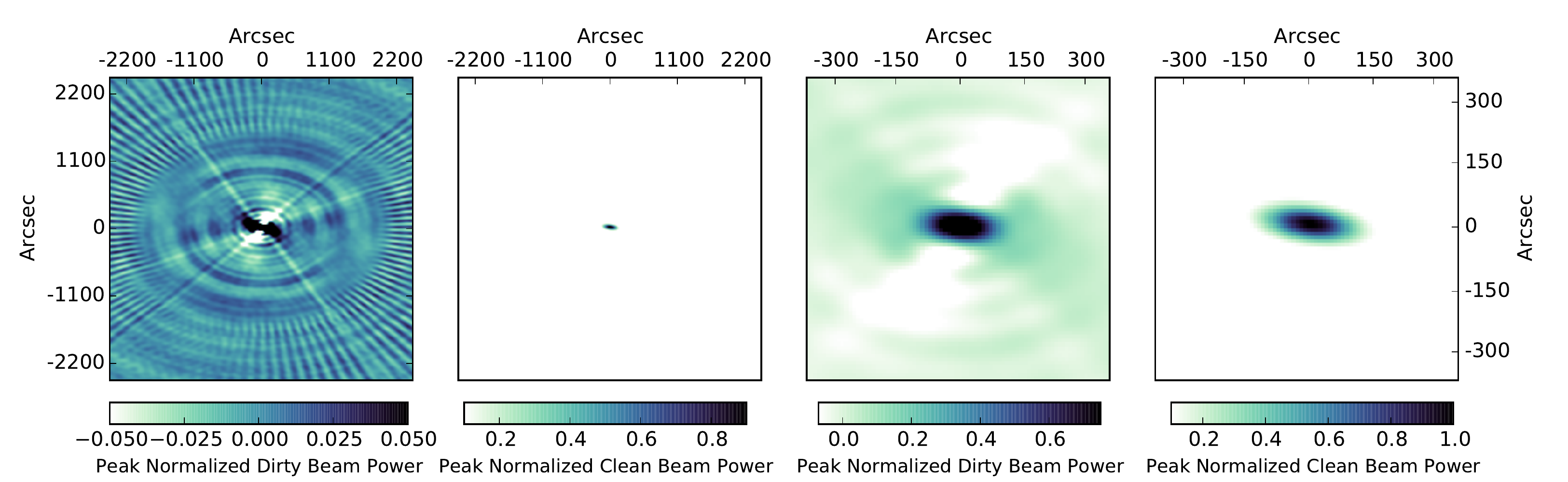}
\caption{Images of the synthesized beams for the $1.75\,$GHz data.. The first two panels are the full `dirty' and `clean' synthesized beams. The third and fourth panels show close ups of the region around the peaks of the beams (dirty and then clean). All beams have been peak-normalized to unity.}
\label{fig:beamim}
\end{figure*}

The goal of this work is to examine large-scale emission, to quantify it, and (if present) to determine what might be causing it. To do this we employ the method of probability of deflection, or {\pd} analysis, also called the $1$-point function \citep[e.g.][]{Scheuer57,Condon74,Patanchon09}. This involves comparing observed and predicted histograms of the image to investigate the underlying source count. Starting from a source count model, a predicted {\pd} can be generated, convolved with noise, and then fit to the observed \pd. A more detailed explanation of {\pd} analysis and derivation of the equations used can be found in the papers cited above or in \citetalias{Vernstrom13}, which we follow in detail. We briefly describe the steps here. 

Starting with a model for the source count, d$N$/d$S$, we compute $R(x)$ as the integral of the count divided by the beam function $B(\theta,\phi)$, which is the beam power pattern peak normalized to unity. $R(x)$ is the mean number of pixels per steradian having {\it observed} intensities between $x$ and d$x$, with $x\equiv SB(\theta,\phi)$, where $S$ is flux density. The predicted {\pd} is then computed from the Fourier transform of $R(x)$, such that

\begin{equation}
\small
P(D) = {\cal{F}}^{-1} \left[ \exp \left( \int_0^{\infty} R\left(x\right) \exp \left(i \omega x\right) dx - \int_0^{\infty} R\left( x \right) dx \right)\right ] \, .
\label{eq:pofw}
\end{equation}
An additional term can be added to account for image noise, as a multiplication in the Fourier domain. This {\pd} calculation assumes only point sources; we discuss the effect of source sizes on the {\pd} in Section~\ref{sec:srcsize}.

In order to fit an accurate {\pd} with a source-count model in this way, the shape of the beam and the image noise must be well understood. Ordinarily one would use a Gaussian model of the synthesized clean beam in the calculation of the model \pd, under the assumption that it is not significantly different from the dirty synthesized beam. However, in our case, the dirty beam has fairly large sidelobes, and is not well approximated by the clean beam. This is shown in Fig.~\ref{fig:beamim}, with the full-sized beams and with a close-up of the regions near the peaks. The peak sidelobes are at about the $\pm 0.1$ level. However, there are pronounced streaks in the outer regions, of amplitude around $\pm0.02$, which, when convolved with a source of $S\approx 100 \, \mu$Jy, would create many pixel values in the $\mu$Jy region. If only the clean beam were used in the calculation then a source count model with a large number of $\mu$Jy sources would be required to achieve a decent fit, even if no such population of sources truly existed. Thus in all following {\pd} calculations we used the dirty beam for all sources below our cleaning limit of $S< 150 \, \mu$Jy, while for sources with $S>150 \, \mu$Jy the clean beam values were used.

\section{Discrete Sources}
\label{sec:discrete}
The discrete source count is now known quite well, and has been shown to provide a very much lower background temperature than the one seen by ARCADE 2, down to at least $50\,$nJy \citepalias{Vernstrom13}. In this paper we are therefore interested in more diffuse extended emission, which would be resolved out at higher resolution. By discrete sources we are referring to sources which are point sources in our $150\,$arcsec$\times60\,$arcsec beam, or sources with $\Omega_{\rm source}\ll \Omega_{\rm beam}$. In order to focus on this emission we first need to subtract out the known contribution from point source emission. We are only able to subtract out sources down to a certain flux density; therefore we must also consider any discrete emission that was not subtracted out.

\subsection{Source subtraction}
\label{sec:subtract}
\begin{figure}
\includegraphics[scale=0.37,natwidth=9in,natheight=9in]{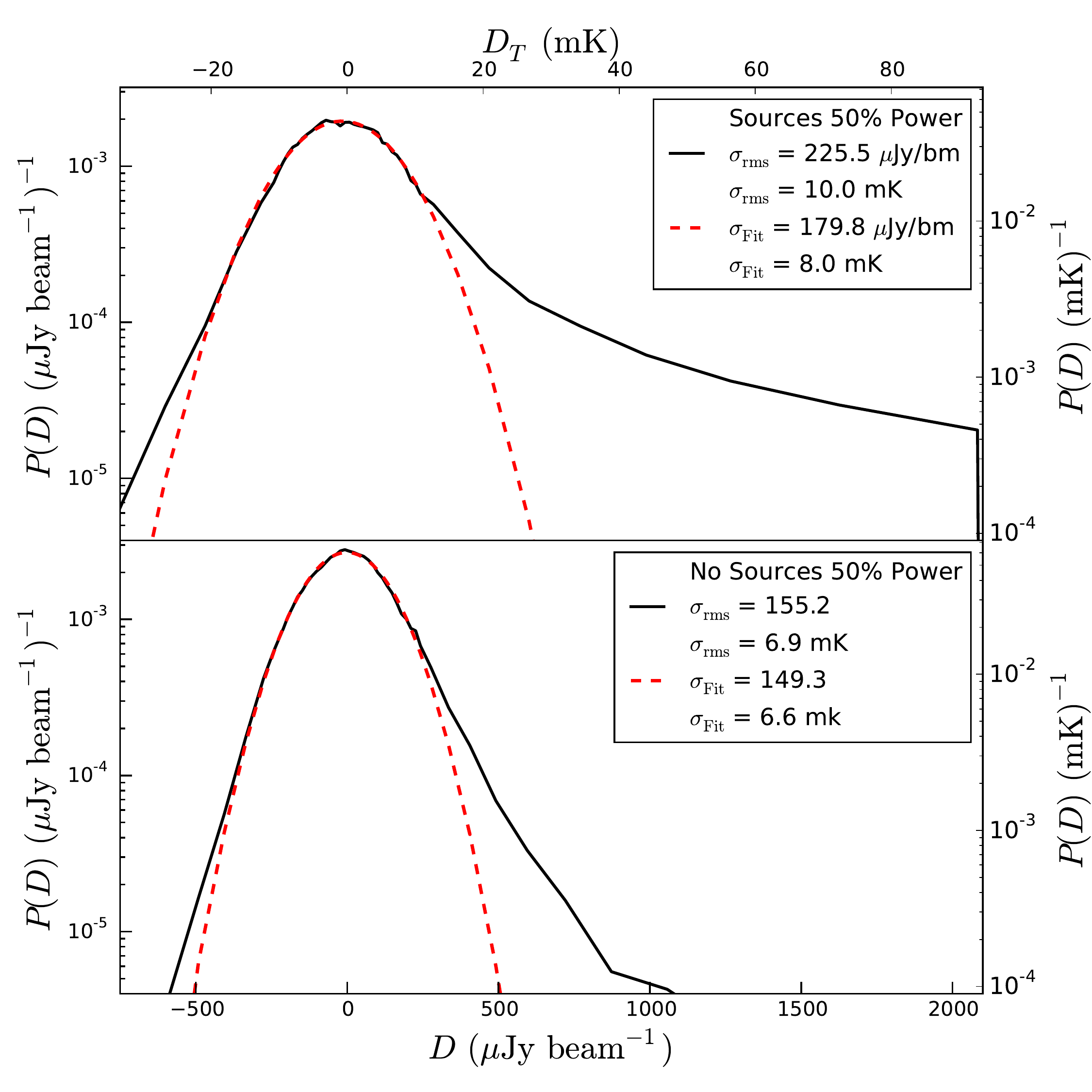}
\caption{{\pd} distributions for the mosaic image central regions, where the increase in noise due to the primary beam is $1.5$ times the minimum noise or lower, an area of roughly $0.61\,$deg$^2$. The top panel shows the pixel histogram for the mosaic before point source subtraction. The bottom panel shows the distribution for the mosaic after subtraction of the ATLAS point sources. The solid black lines are the image distributions, while the red dashed lines are fitted Gaussians.}
\label{fig:mospdf}
\end{figure}

We used the clean component models from the ATLAS survey third data release (Franzen et. al, 2014 in preparation, Banfield et. al, 2014 in preparation) as point source models for subtraction, since the ATLAS resolution is significantly higher than our data, at around $10\,$arcsec. It is not entirely clear what the median source size might be and how it changes with flux density, but we expect a value between 1 and $3\,$arcsec for typical galaxies in evolutionary models \citep[e.g.][]{Wilman08}. Thus the ATLAS resolution should be sufficient to measure all of the discrete or point source emission. The ATLAS point source models were split into two frequency bands: the lower frequencies from $1.30$ to $1.48\,$GHz; and the higher frequencies from $1.63$ to $1.80\,$GHz. For the subtraction we split our seven \textit{uv}-data sets (for each pointing) into two equal frequency bands as well: $1.30$ to $1.70\,$GHz; and $1.70$ to $2.10\,$GHz. The ATLAS images were made using multi-frequency deconvolution and thus contain estimates of the spectral indices of the clean components, which can be used to scale the flux density to different frequencies during subtraction. The task \textsc{uvmodel} was used to subtract the appropriate pointing and frequency coverage clean model from each corresponding \textit{uv}-data set; then the \textit{uv}-data for each pointing were concatenated using the task \textsc{uvglue} (combining the lower and upper frequency parts for each pointing). An independent image was constructed from each pointing with a mosaic constructed subsequently. 

The ATLAS survey has an rms sensitivity of $15$ to $25\, \mu$Jy beam$^{-1}$ (depending on the pointing) and the models were cleaned down to a level of $150\, {\rm \mu}$Jy beam$^{-1}$. Thus all point sources with $S>150\, {\rm \mu}$Jy should have some fraction of their discrete emission subtracted out. There is some residual emission apparent around the brightest sources, which is visible in the bottom right panel of Fig.~\ref{fig:images}. We cannot say if this is due to some slight calibration or subtraction error, possible time variability of AGN sources, or if this represents a portion of the sources' diffuse emission. Looking at the peak positions of the well defined objects in each of the images, the average residual is only $5\,$per cent of the peaks. The \pd s for the central region of the mosaic images before and after source subtraction are presented in Fig.~\ref{fig:mospdf}; this shows a clear decrease in the size of the positive source tail for the subtracted image. 

When comparing our data to {\pd} predictions from source-count models we use the {\pd} of the source-subtracted mosaic image, including only pixels from regions where the noise due to the primary beam correction is not more than 1.5 times the lowest noise value. This is because the {\pd} calculation from a source-count model assumes a constant value for image noise. The noise is certainly inhomogeneous in our data. However, simulations have shown that the effect on the {\pd} calculation is small if we limit ourselves to a region where the change in the noise is small and create a noise-weighted histogram. Using a weighting scheme described in \citetalias{Vernstrom13}, we calculate a mean noise in this area (approximately $0.61\,$deg$^2$) of $\sigma_{\rm n}=(52\pm5)\, \mu$Jy beam$^{-1}$, or ($2.3\pm0.2\,$mK).

\subsection{Counts and confusion}
\label{sec:ccounts}

\begin{figure}
\includegraphics[scale=0.37,natwidth=9in,natheight=11in]{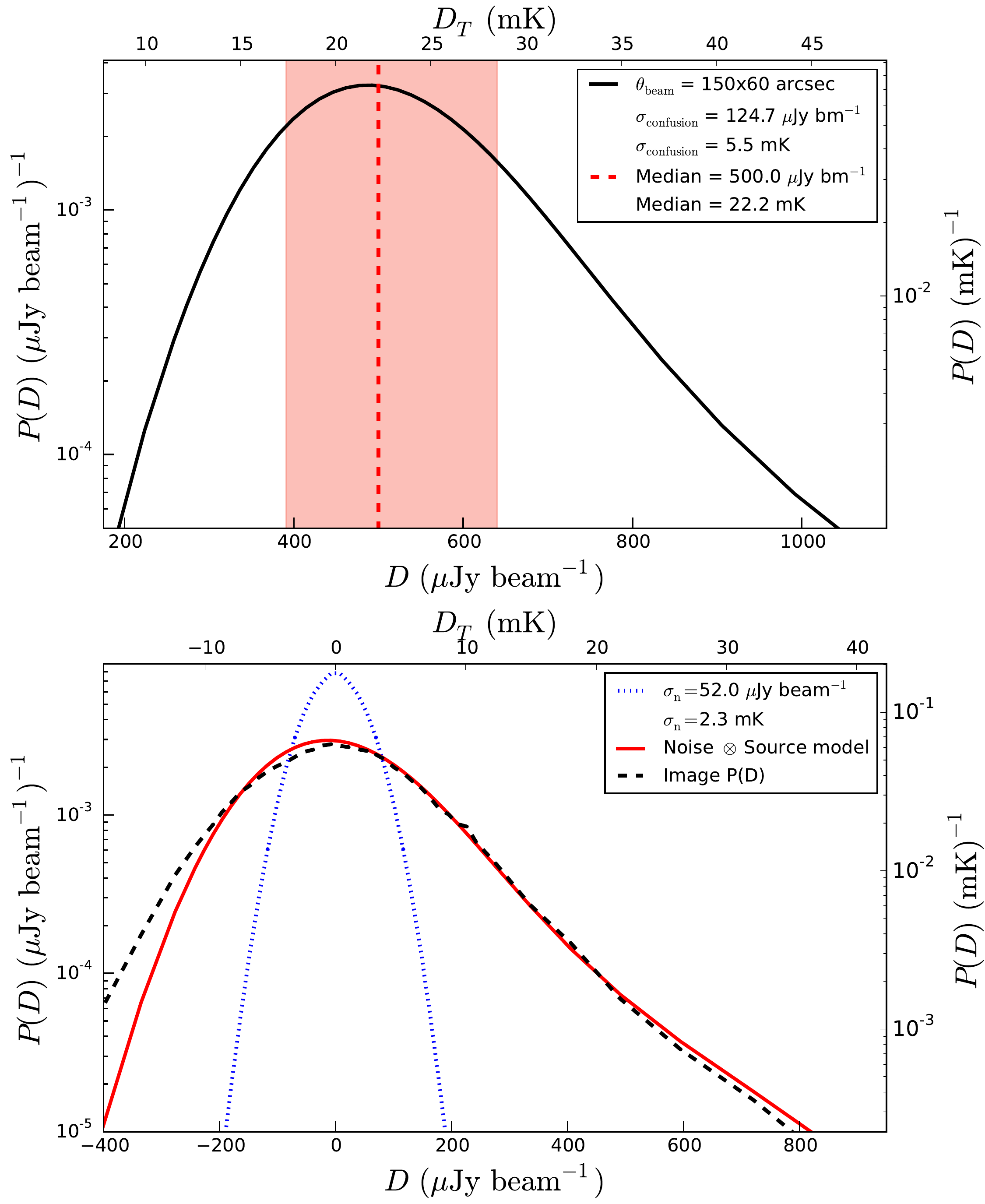}
\caption{Source confusion distribution for discrete sources (point sources in our $150\,$arcsec$\times60\,$arcsec beam). The top panel shows the noiseless {\pd} from the source count of \citetalias{Vernstrom13}, scaled from $3\,$GHz to $1.75\,$GHz using $\alpha=-0.7$, including only sources up to a flux density of $S=150 \, {\rm \mu}$Jy, with a differential source count logarithmic slope of $-2.5$ for $150\le S \le 3000 \, \mu$Jy. The measured confusion rms from this distribution is $\sigma_{\rm c}= (125 \pm 10) \, {\rm \mu}$Jy beam$^{-1}$, or $(5.5\pm0.44)\,$mK, with the dashed lines showing the median and the shaded regiom showing the $\pm1\sigma$ values. The bottom panel shows this distribution convolved with a Gaussian of width $\sigma_{\rm n}=52\, {\rm \mu}$Jy beam$^{-1}$. The Gaussian is the blue dotted line, the convolution is the red solid line, and the ${\pd}$ from the inner region of the source-subtracted mosaic image is shown as the black dashed line. }
\label{fig:conf}
\end{figure}

We need to estimate the contribution of discrete emission from sources that were not subtracted out. For sources below the clean threshold of the ATLAS models we take the discrete source count of \citetalias{Vernstrom13}, including sources up to $S=150 \, {\rm \mu}$Jy, which is measured via confusion analysis down to $S\simeq0.05 \, {\rm \mu}$Jy at $3\,$GHz. We scale this to $1.75\,$GHz according to $S \propto \nu^{\alpha}$, with $\alpha=-0.70 \pm 0.05$ being the mean spectral index of star-forming galaxies \citep{Condon84b}. We found that slight variation in this spectral index produces no significant effect on the output. 

For the bright residuals left over from the subtraction process the issue is not as straightforward. Even neglecting any errors in calibration or subtraction, the clean process which generated the models is highly non-linear. The clean components may only represent a fraction of the true flux density, which can vary by peak flux density and from pointing to pointing. We do not believe there to be extended emission brighter than approximately $150\, \mu$Jy beam$^{-1}$ (as discussed in more detail in Section~\ref{sec:highres}). To account for unresolved residuals brighter than this we counted all the peaks in the source-subtracted image brighter than $150\, \mu$Jy beam$^{-1}$ that are associated with point sources in the image with no subtraction, and calculated a power law index for their differential source count of $-2.50$. 

Our model for the unsubtracted point-source contribution is then the scaled \citetalias{Vernstrom13} source count up to $150\, \mu$Jy with a power law of slope $-2.50$ attached for sources with $150 \, \mu {\rm Jy} < S < 3\,$mJy ($3\,$mJy being the brightest residual in the fitting area). We computed the {\pd} from this count and convolved this {\pd} with a Gaussian noise distribution of width $\sigma_{\rm n}=(52\pm5) \, {\rm \mu}$Jy beam$^{-1}$, or $(2.3\pm0.2)\,$mK. The noiseless and convolved {\pd} distributions are shown in Fig.~\ref{fig:conf}. We measured the confusion noise $\sigma_{\rm c}$, or width of the distribution, by first finding $D_1$ and $D_2$,
\begin{equation}
 \sum_{D_1}^{\rm median} P(D) =\sum_{\rm median}^{D_2} P(D) =0.34,
 \label{eq:confuse}
 \end{equation}
 when normalised such that the sum over the $P(D)$ is 1. Then we take $\sigma_{\rm c}=(D_2-D_1)/2$. We do this  since, in the Gaussian case, $68\,$per cent of the area is between $\pm 1\sigma$, and since, in the more realistic case, the long positive tail makes the variance of the full distribution a poor estimate of the width if the peak. The estimated width of the source-subtracted image {\pd} is $\sigma=155 \, \mu$Jy beam$^{-1}$ ($\,6.9\,$mK) with an uncertainty of $\pm 5\, \mu$Jy beam$^{-1}$ ($\pm0.22\,$mK) measured from bootstrap resampling. For the discrete source model {\pd} we find a value of $\sigma_{\rm c}=125 \, {\rm \mu}$Jy beam$^{-1}$ ($\,5.5\,$mK). The {\pd} of this model  convolved with Gaussian noise thus has an rms of $\sigma_{\rm c \otimes n}=135 \, \mu$Jy beam$^{-1}$ ($\, 6.0\,$mK). 

This discrete model estimate should be treated with some caution. The result is dependent on the exact value of the noise used in the calculation and the exact shape of the unsubtracted discrete count contribution. The unsubtracted discrete count is based on a model which is dependent on the maximum flux density value for the point sources with no subtraction, as well as the power law used for the brighter sources. Taking these points into consideration we adopt an uncertainty of $\pm 10\ \, \mu$Jy beam$^{-1}$, or $\,\pm0.44\,$mK, on the measure of  $\sigma_{\rm c}=125 \, \mu$Jy beam$^{-1}$=$\,5.5\,$mK, yielding a measurement and uncertainty for the width of the noise convolved distribution of $\sigma_{\rm c \otimes n}=(135 \pm 12) \, \mu$Jy beam$^{-1}$, or $(6.0\pm0.53)\,$mK.
 
We want to know how different the model of unsubtracted discrete source emission is from the data. To do this we performed a bootstrap significance test. We randomly selected a subset of half the image pixels, generated random numbers from the noiseless model distribution and added varying amounts of Gaussian noise (to account for the uncertainty in the model). We then combined the real and model data into one set and drew two new subsets at random from the combined distribution. We compared the binned real data to the binned model data, and the binned combined random sets to each other. We repeated this procedure 5000 times. This yields a distribution of the test statistic from the combined random samples of the null hypothesis (that the observed and model data come from the same population) and a distribution of the test statistic when comparing the ordered sets (the observed and model sets not combined). We computed three different test statistics: the Euclidean distance (the root-mean-square distance between the histograms); the Jeffries-Matusita distance (similar to the Euclidean distance but more sensitive to differences in small number bins); and a simple $\chi^2$. 

The results of the bootstrap test show an average excess width of $(76 \pm 23)\, \mu$Jy beam$^{-1}$, $(3.4\pm1.0)\,$mK, with the value of $76$ coming from $\sqrt{\sigma^2- \sigma_{\rm c \otimes n}^2}=\sqrt{155^2-135^2}$. The exact significance of this excess varies depending on the test statistic . However, regardless of which test statistic is used the data and model are statistically different, with a minimum of $99.5\,$per cent confidence. This excess cannot be converted directly into a background temperature since the conversion depends on the underlying source-count model responsible for the width (see Section~\ref{sec:btemp} for more discussion on the temperature conversion).

Based on these tests, we conclude that there is more emission present than that from compact galaxies alone at the roughly $3\sigma$ level. However, due to the uncertainty in the source subtraction process, this excess and any resulting extended emission models are here considered as upper limits on the extended emission present.
 
\section{Extended Sources}
\label{sec:extcnt}

\subsection{High resolution imaging of extended emission}
\label{sec:highres}

Before attempting to model any extended emission in the ATCA data we consider how extended emission is detected at higher resolutions, comparing the VLA data used by \citetalias{Vernstrom13} and the ATLAS ATCA high resolution images. The VLA $3$-GHz beam used in \citetalias{Vernstrom13} had a FWHM of $8\,$arcsec, while the ATLAS beam was roughly $10\,$arcsec. We would like to know how emission on arcmin scales appears with these types of observations, since we know that some emission will be resolved out at higher resolution. 

This was tested using sources from $1\,$deg$^2$ of the SKADS simulation \citep{Wilman08} at $1.4\,$GHz. This simulation was shown in \citetalias{Vernstrom13} to be a close approximation to observed source counts. Using the flux densities provided, we made one image containing only point sources. Then assuming each point source has an extended halo with total flux set to $S_{\rm dis}/10$, with $S_{\rm dis}$ being the point source (discrete) flux, we made two images, assuming all the haloes were Gaussians with FWHM of $30$ or $60\,$arcsec. We added the point sources to these and convolved the images with a $9\,$arcsec beam (the average size of the VLA and ATCA resolutions). 

\begin{figure}
\centering
\includegraphics[scale=0.32,natwidth=11in,natheight=8.5in]{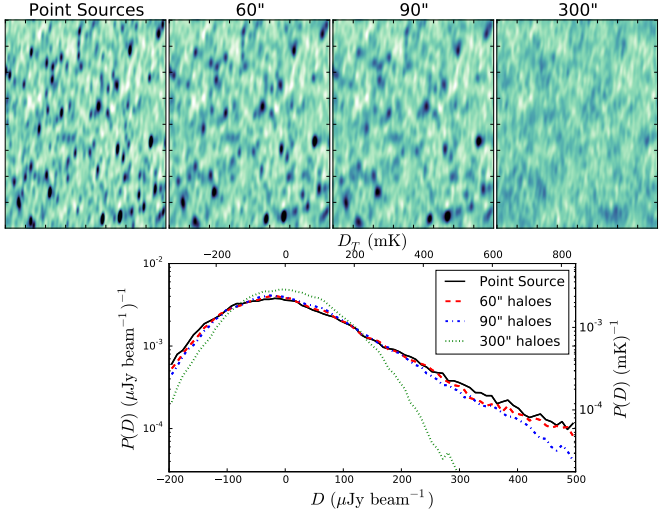}
\caption{Simulation showing point source and extended emission at higher resolution. The top panels show cut-outs of the simulation with just point source emission (left), point sources plus haloes of $30\,$arcsec diameter (middle), and point sources plus haloes of $60\,$arcsec diameter (right), all convolved with a $9\,$arcsec beam and with Gaussian noise of $2\, \mu$Jy beam$^{-1}$. The total flux density of each halo is taken as the point source flux density divided by 10. The bottom panel shows the {\pd} distributions from the three images, with the solid black line being for point sources only, the red dashed line point sources plus $30\,$arcsec haloes, and the blue dot-dashed line point sources plus $60\,$arcsec haloes. }
\label{fig:simims}
\end{figure}

The confusion noise of each of the noiseless images are $1.53,1.95,$ and $1.78\, \mu$Jy beam$^{-1}$ for the discrete, discrete$+30\,$arcsec, and discrete$+60\,$arcsec data sets at $1.4\,$GHz. The $30\,$arcsec haloes add a width of $\sigma_{30}=\sqrt{1.95^2-1.53^2}=1.21\, \mu$Jy beam$^{-1}$, and the $60\,$arcsec haloes add $\sigma_{60}=\sqrt{1.78^2-1.53^2}=0.91\, \mu$Jy beam$^{-1}$. The \pd s for the images with point sources plus extended emission are shown in the bottom panel of Fig.~\ref{fig:simims}. The smaller the extended objects the greater the increase in the width of the distribution. For images with the same total flux density the distribution for the larger sources would have its DC level shifted to higher flux densities; however interferometers are not sensitive to the DC level (or lowest spatial frequency) and thus do not measure total flux densities.  

The measured confusion rms from the $3\,$GHz VLA data is approximately $(1.2 \pm 0.07) \, \mu$Jy beam$^{-1}$ (depending on the source-count model). Scaling the simulated values to $3\,$GHz, the addition of the $30\,$arcsec extended emission to the VLA point source model would yield a width of $1.38\, \mu$Jy beam$^{-1}$, with the $60\,$arcsec haloes yielding $1.31\, \mu$Jy beam$^{-1}$. Although in this case these exceed the estimated uncertainty, the simulated confusion widths depend on the exact source count used and the assumption of how the extended emission depends on the point-source flux density. Thus these particular extended emission models produce excess widths in the {\pd} distributions that are large enough to have been detected in deep high resolution images. However, these simulations show that there likely exist models with either fainter or larger-scale extended emission that would have been undetected in the VLA {\pd} experiment of \citetalias{Vernstrom13}.  

From the simulated extended-size images we see that sources with total halo flux densities greater than approximately $150 \, \mu$Jy would be visible in the VLA or ATLAS images. The top panel of Fig.~\ref{fig:simims} shows a cut-out of the simulated images; when the point-source flux density is faint ($\leq 200 \, \mu$Jy), the extended emission is not visible in the image. However, for brighter point sources (with brighter halo emission), the extended haloes are visible in both the $30$ and $60\,$arcsec images. Since nothing of this nature is seen in either the VLA or ATLAS images, we conclude that any extended emission in the current low resolution ATCA data should also have total flux density less than about $150 \, \mu$Jy, or else has to be very rare.  

\subsection{Extend source size sensitivity}
\label{sec:srcsize}
The {\pd} calculation does not use any size information and assumes only unresolved sources. Therefore, it is important to understand how resolution affects the {\pd} fitting. To test this we used the simulated halo flux densities for the extended emission described in Section~\ref{sec:highres} and made four separate images for sources treated simply as point sources and as Gaussians with FWHM of 60, 90, and $300\,$ arcsec. These give a range of sizes in relation to the ATCA beam. We then ran each image through the fitting routine for source-count amplitudes at specific flux densities, i.e. a set of connected power laws \citep[e.g.][]{Patanchon09,Vernstrom13}. 

\begin{figure}
\includegraphics[scale=0.37,natwidth=9in,natheight=9in]{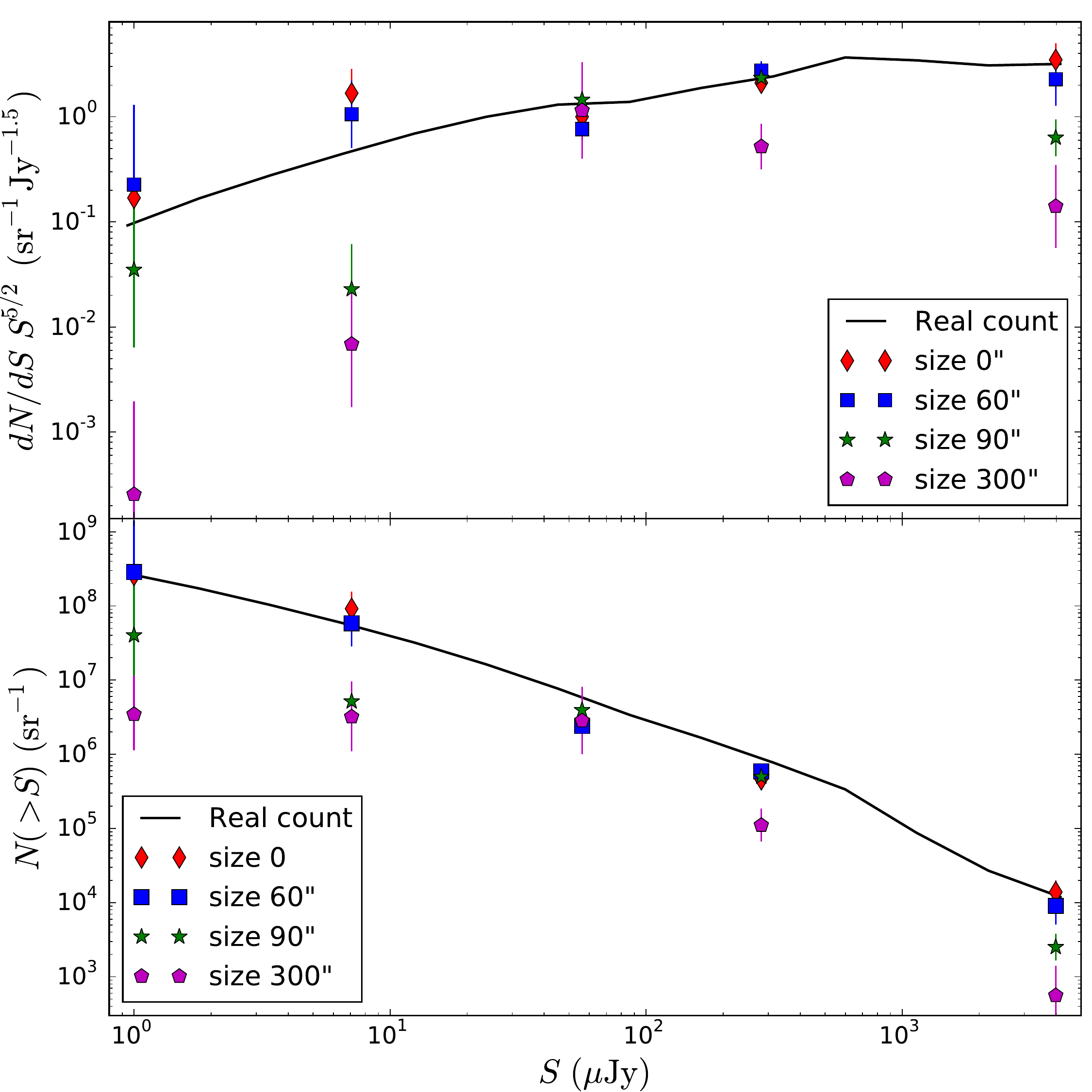}
\caption{Results of {\pd} fitting of simulated images with different source sizes. The top panel shows the Euclidean-normalized differential source count of the input count (solid black line) and best-fitting results of the point source image (red diamonds), the $60\,$arcsec size image (blue squares), the $90\,$arcsec size image (green stars), and the $300\,$arcsec size image (magenta pentagons). The bottom panel shows the same information, but plotted as integrated source counts.}
\label{fig:simvlact}
\end{figure}

The results show that there is no significant change in the fitting results between the point source and $60\,$arcsec size images. However, the results for the $90\,$arcsec sizes are lower at both the faintest and brightest flux densities, while the $300\,$arcsec size results are significantly lower at all flux densities. The results of this test are presented in Fig.~\ref{fig:simvlact}. This shows that the {\pd} fitting procedure is reliable for sources on the order of the beam size or smaller; Table~\ref{tab:zsizes} shows the linear sizes for the angular scales to which we are sensitive given a range of redshifts, assuming standard $\Lambda$CDM cosmology with $H_{0}=67.8\,$km s$^{-1}$ Mpc$^{-1}$, $\Omega_{\rm m}=0.308$, and $\Omega_{\Lambda}=0.692$ \citep{Planck13a}. 

\begin{table}
 \centering
  \caption{Angular and physical source sizes at different redshifts.}
  \begin{tabular}{ccccc}
\hline 
\hline
 Angular Size & \multicolumn{4}{c}{Physical Size}\\
 & $z=0.25$ & $z=0.5$ & $z=1$ & $z=2$\\
(arcsec) & (Mpc) & (Mpc)& (Mpc) & (Mpc)\\
\hline
$30$&$0.12$&$0.19$&$0.25$&$0.26$\\
$60$&$0.24$&$0.38$&$0.49$&$0.51$\\
$100$&$0.40$&$0.63$&$0.82$&$0.86$\\
$150$&$0.60$&$0.94$&$1.23$&$1.29$\\
\hline
\end{tabular}
\label{tab:zsizes}
\end{table}

\section{Extended Source Count Models}
\label{sec:mods}
We have shown that there is a significant excess in the width of the observed distribution over that estimated from noise and discrete point sources, suggesting the presence of diffuse or extended sources. This emission could be low surface brightness diffuse emission around individual galaxies, diffuse cluster emission, or something more exotic, such as emission from dark matter annihilation in haloes. We now use three source count models to investigate the possible excess (extended) emission. We follow the fitting procedure described in detail in \citetalias{Vernstrom13}. We use Monte Carlo Markov Chains (MCMC), employing the software package \textsc{CosmoMC} \citep{Lewis02}\footnote{\url{http://cosmologist.info/cosmomc/}}, to minimize $\chi^2$ for each model. The three most negative bins ($-500 \le D (\, \mu{\rm Jy} \,{\rm beam}^{-1}) \le -250$) from the image histogram were neglected in the calculation of $\chi^2$. This is because the data have a clearly non-Gaussian negative tail, due in part to the non-constant noise but also due to the areas of over-subtraction, which produce an excess of negative points (see the bottom panel of Fig.~\ref{fig:mospdf}). Tests on subsets of the data, and using different detailed approaches for subtracting bright sources, showed that these effects were restricted to the most negative bins, with the rest of the histogram being quite stable. 

\subsection{Shifted discrete count model}
\label{sec:dshifts}

\begin{figure}
\includegraphics[scale=0.37,natwidth=9in,natheight=9in]{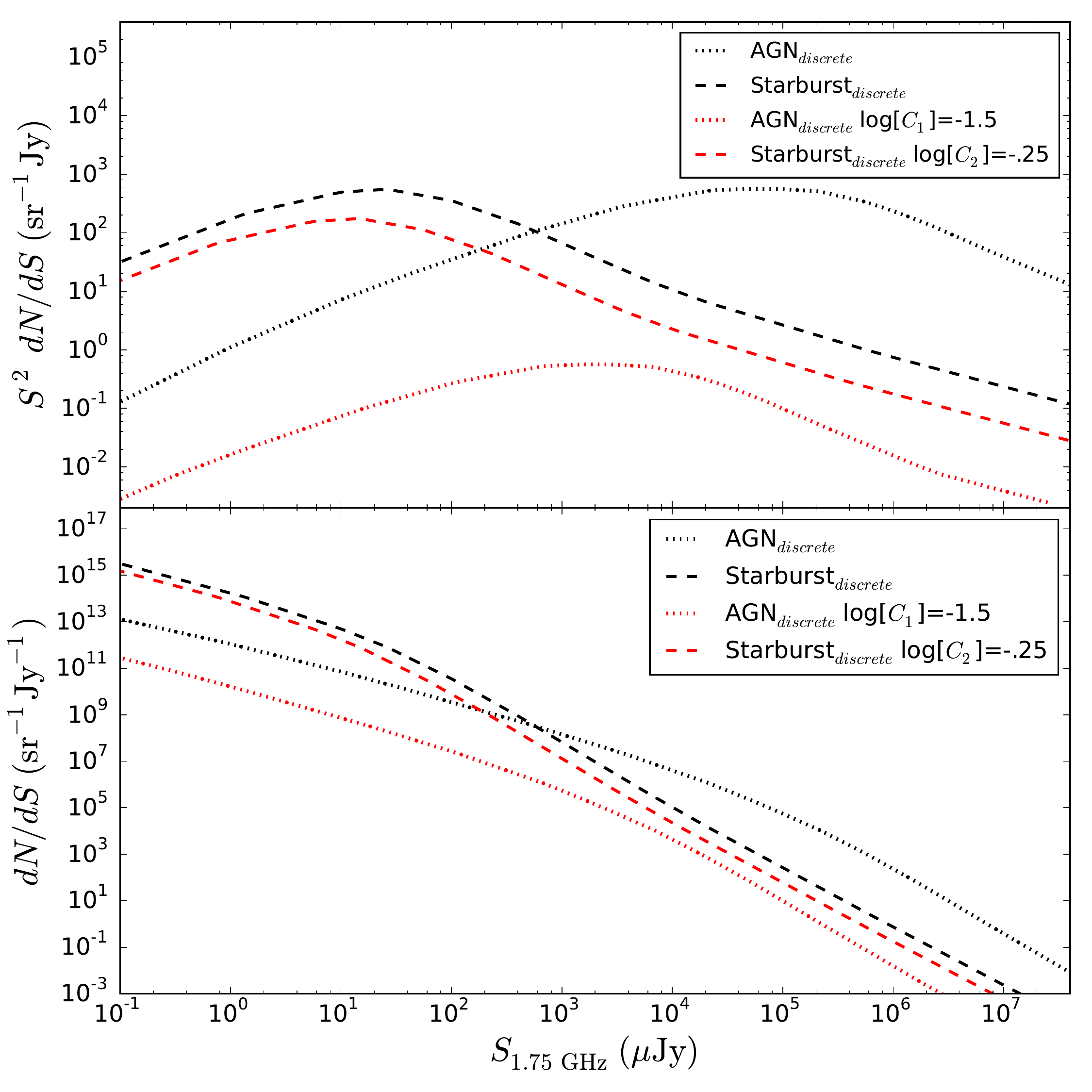}
\caption{Discrete and shifted source counts of AGN and starburst. The top panel shows the discrete AGN and starburst source counts (black dotted line and black dashed line) using $S^2$ normalization. The red lines are example of the shifting model described in Section~\ref{sec:dshifts}, where $S$ for the AGN count has been shifted by $\log_{10}[C_1]=-1.5$ and the starburst count is shifted by $\log_{10}[C_2]=-0.25$. The bottom panel shows the same lines with {\it no} normalization on the source counts. This demonstrates how applying only a horizontal shift in $\log_{10}[S]$ will appear as a combination of vertical or amplitude shift when the $S^2$ normalization is applied.  }
\label{fig:shiftex}
\end{figure}

Using evolutionary models \citep[e.g.][]{Condon84b,Hopkins00} the source count can be broken into contributions from two populations, namely AGN and star-forming galaxies, as shown in Fig.~\ref{fig:shiftex}. The simplest extended-emission model assumes that each of these populations has a radio-emitting halo on arcmin scales, proportional to some fraction of the discrete flux density (or $S_{\rm discrete}\times C$), separately for the two populations. The counts associated with this extended emission must then retain the shape of the discrete counts for each population, but can be shifted in flux density. To estimate the extended counts that are consistent with our data we took the discrete counts for each population and simply applied a shift in $\log_{10}[S]$ separately. Thus,
\begin{equation}
\begin{split}
\frac{{\rm d}N(S_{\rm ext})_{\rm AGN}}{{\rm d}S_{\rm ext}}=\frac{{\rm d}N([S_{\rm dis}C_1])_{\rm AGN}}{{\rm d}[S_{\rm dis}C_1]},\\
\frac{{\rm d}N(S_{\rm ext})_{\rm SB}}{{\rm d}S_{\rm ext}}=\frac{{\rm d}N([S_{\rm dis}C_2])_{\rm SB}}{{\rm d}[S_{\rm dis}C_2]},
\label{eq:ns}
\end{split}
\end{equation}
where $C_{1}$ and $C_{2}$ are constants that are varied to fit the counts. When combined with the unsubtracted discrete count and Gaussian noise, we can find the values that best fit the observed {\pd} distribution of our source-subtracted image. Figure~\ref{fig:shiftex} shows an example of this model with the two populations of discrete counts, each with shifts applied. We plot the results with the usual $S^2$ normalization and with no normalization (so that the horizontal shifts can be seen).This model will be referred to as Model 1.

\subsection{Parabola model}
\label{sec:pop3}
We would also like to investigate the possibility of the extra emission being fit by a single new population. To do this we introduce a new population as a parabola in $\log_{10}[S^2{\rm d}N/{\rm d}S]$ of the form
\begin{equation}
S^2\frac{{\rm d}N(S)_{\rm ext}}{{\rm d}S}=A(x-h)^2+k.
\label{eq:quad}
\end{equation}
Here $x=\log_{10}[S]$ and $A,\,h,$ and $k$ are all free parameters. The parameter $h$ is the peak position in $\log_{10}[S]$, $k$ is the amplitude or height of the peak, and $A$ (along with $k$) controls the width. We chose this model because it allows for a smooth curve, and since the discrete count populations are themselves crudely approximated by parabolas in $\log_{10}[S^2{\rm d}N/{\rm d}S]$. This model will be referred to as Model 2.

\subsection{Node model}
\label{sec:nodemod}
There may be several types of sources or populations contributing to the extended emission counts, including individual galaxies, clusters, dark matter, intra-cluster medium, etc. Without having physical models for these different populations, we would require too many parameters to fit separate models for each. Therefore, we have chosen also to fit a model of connected power laws. This model allows for the shape of the source count to vary over a particular flux density range, rather than having a fixed shape based on a few parameters. It therefore has the potential to be sensitive to contributions from different populations at different flux densities.

The model consists of fitting for the amplitude of $\log_{10}[dN/dS]$ at specific flux densities, or nodes, and interpolating linearly (in log space) between the nodes -- for more details on this model see \citetalias{Vernstrom13}. We specifically use five nodes spaced evenly in $\log_{10}[S]$, covering the range of $0.5 \le S \le 1000 \, \mu$Jy.  This model will be referred to as Model 3.

\section{Background Temperature}
\label{sec:btemp}
\subsection{Source count contribution to temperature}
\label{sec:sc2t}
 
We can easily obtain an estimate of the contribution from sources to the radio background temperature. The source count and the sky temperature at a frequency $\nu$ are related by the Rayleigh-Jeans approximation,
\begin{equation}
\int_{S_{\rm min}}^\infty S \frac{dN}{dS} dS=  {T_{\rm b} 2 k_{\rm B}\nu^2 \over  c^2 }. 
\label{eq:tint}
\end{equation}
In the above equation $k_{\rm B}$ is the Boltzmann constant, and $T_{\rm b}$ is the sky temperature from all the sources brighter than $S_{\rm min}$. Equation~(\ref{eq:tint}) is also equivalent to
\begin{displaymath}
\int_{S_{\rm min}}^\infty S^2 \frac{dN}{dS} d[\ln(S)]=  {T_{\rm b} 2 k_{\rm B}\nu^2  \over  c^2 }. 
\end{displaymath}
It is for this reason that it is convenient to show the source count weighted not by the Euclidean $S^{5/2}$ but by $S^{2}$, e.g. as in Fig.~\ref{fig:shiftex}. This alternate weighting of $S^{2}dN/dS$ is proportional to the source count contribution to the background temperature per decade of flux density. With such a plot the source count must fall off at both ends to avoid over-predicting the background (i.e. violating Olbers paradox); hence the bright end must turn over at flux densities above those we have plotted. The discrete source count used by \citetalias{Vernstrom13} integrates (up to $S=900\,$Jy) to a background temperature at $1.75\,$GHz of $T_{\rm dis}(1.75\,{\rm GHz})=63\,$mK, where $T_{\rm dis}$ is the temperature from the discrete source contribution.

\subsection{ARCADE 2}
\label{sec:arctemp}
The ARCADE 2 experiment measured a background temperature of $(54\pm6) \,$mK at $3.3\,$GHz. Several fits are provided to these data, which allow for scaling of the result to different frequencies. The initial fit provided in \citet{Seiffert09} is
\begin{equation}
T_{\rm b}=(1.06 \pm 0.11 {\rm K})\left (\frac{\nu}{1 {\rm GHz}} \right )^{-2.56\pm0.04}.
\label{eq:arct1}
\end{equation}
There is another fit, incorporating data from lower frequencies, given in \citet{Fixsen11} as
\begin{equation}
T_{\rm b}=(24.1\pm2.1 {\rm K})\left (\frac{\nu}{310 {\rm MHz}} \right )^{-2.599\pm0.036}.
\label{eq:arct2}
\end{equation}
Using both of these fits we calculated the estimated background temperature at $1.75\,$GHz by taking the average from the two equations, and an uncertainty using the highest and lowest values from the uncertainties in the equation parameters. This yields $T_{\rm AR2}(1.75\, {\rm GHz})=(265 \pm 45)\,$mK, which corresponds to a total flux density, given our beam size, of $5600 \, \mu$Jy beam$^{-1}$. 

In addition to fitting the data with no constraints, we also fit the models to see what kind of count shapes would be necessary to achieve the ARCADE 2 temperature. We fit the models as described above, only this time adding a prior requiring that the integrated temperature be in the range of 150 to $300\,$mK. This should show if there is any such source count model consistent with both ARCADE 2 and our data. 

These models are referred to as Model 1A (shifts), Model 2A (parabola), and Model 3A (nodes).

\section{Results}
\label{sec:results}
\subsection{Summary of fitting results}
\label{sec:res1}
Using the three models from Section~\ref{sec:mods} we (a) examined what model parameters best fit our new ATCA data, (b) calculated the resulting contribution to the background brightness temperature, and (c) modelled what would be necessary to achieve a background temperature consistent with ARCADE 2. 

\begin{figure}
\includegraphics[scale=0.37,natwidth=9in,natheight=9in]{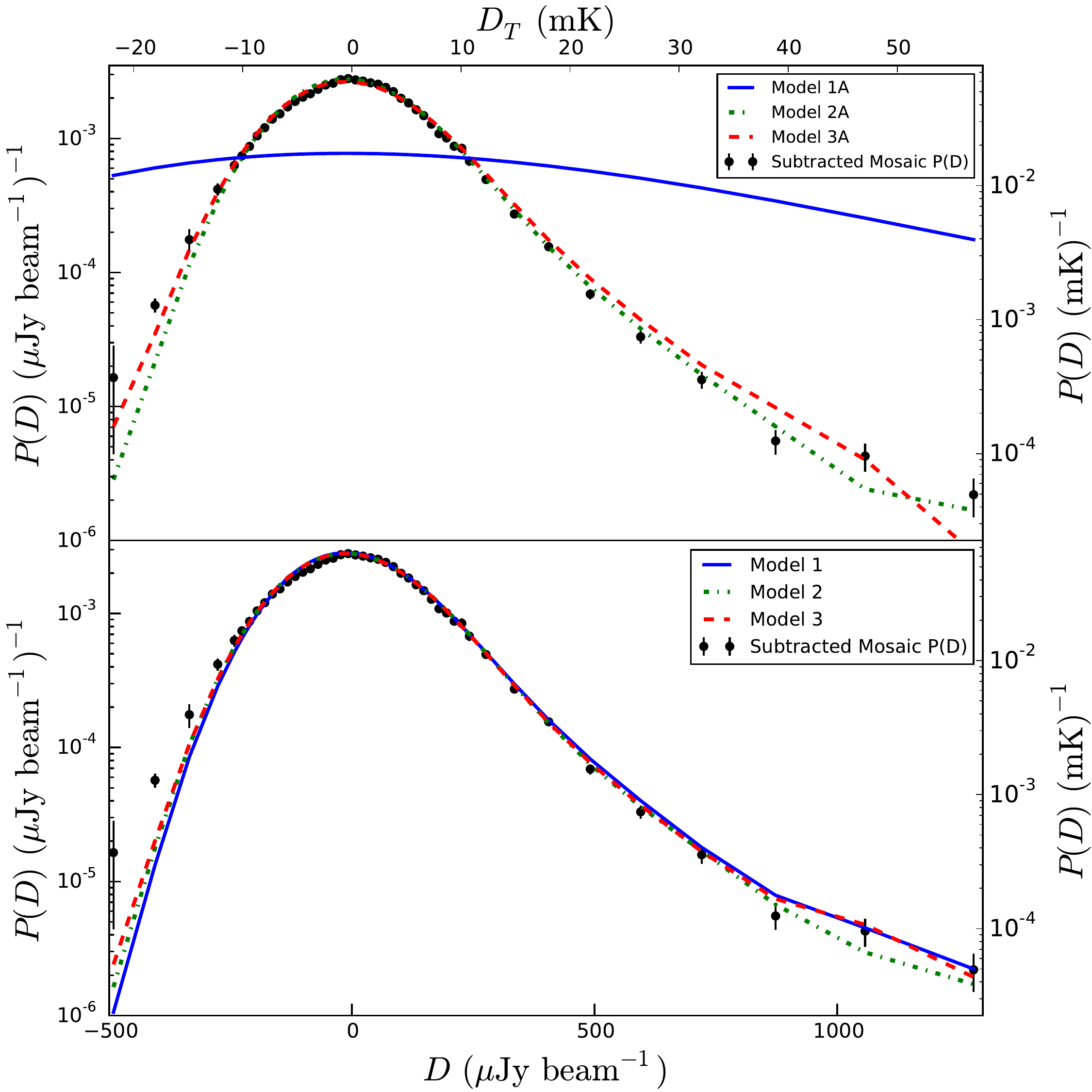}
\caption{{\pd} distributions for various extended-emission source counts. The top panel shows the {\pd} distributions for the best-fitting models of extended-emission counts with a prior for the ARCADE 2 temperature for Model 1A (blue solid line), Model 2A (green dot-dashed line) , and Model 3A (red dashed line). The bottom panel shows the results of fitting the same models, but without the temperature requirement. All models have been convolved with Gaussian noise of $\sigma_{\rm n}=52\,  {\rm {\mu}}$Jy beam$^{-1}$ and the unsubtracted discrete source count contribution. The black points are the source-subtracted mosaic histogram (as seen in the bottom of Fig.~\ref{fig:mospdf}). }
\label{fig:temps}
\end{figure}

The results from Model 1 and Model 1A are in Table~\ref{tab:shifts}, results from fitting Model 2 and 2A are in Table~\ref{tab:p3pop}, and results from Model 3 and Model 3A are listed in Table~\ref{tab:pmnode}.  Each of these extended counts was added to the unsubtracted discrete count model (discrete source count fainter than the subtraction limit plus a power law for subtraction residuals, as discussed in Section~\ref{sec:ccounts}) to compute the {\pd} for each model. The {\pd} models, convolved with Gaussian noise of $52\, \mu$Jy beam$^{-1}$, are shown in Fig.~\ref{fig:temps}, along with the {\pd} for the central region of our source-subtracted mosaic image. 

\begin{figure}
\includegraphics[scale=0.35,natwidth=9in,natheight=12in]{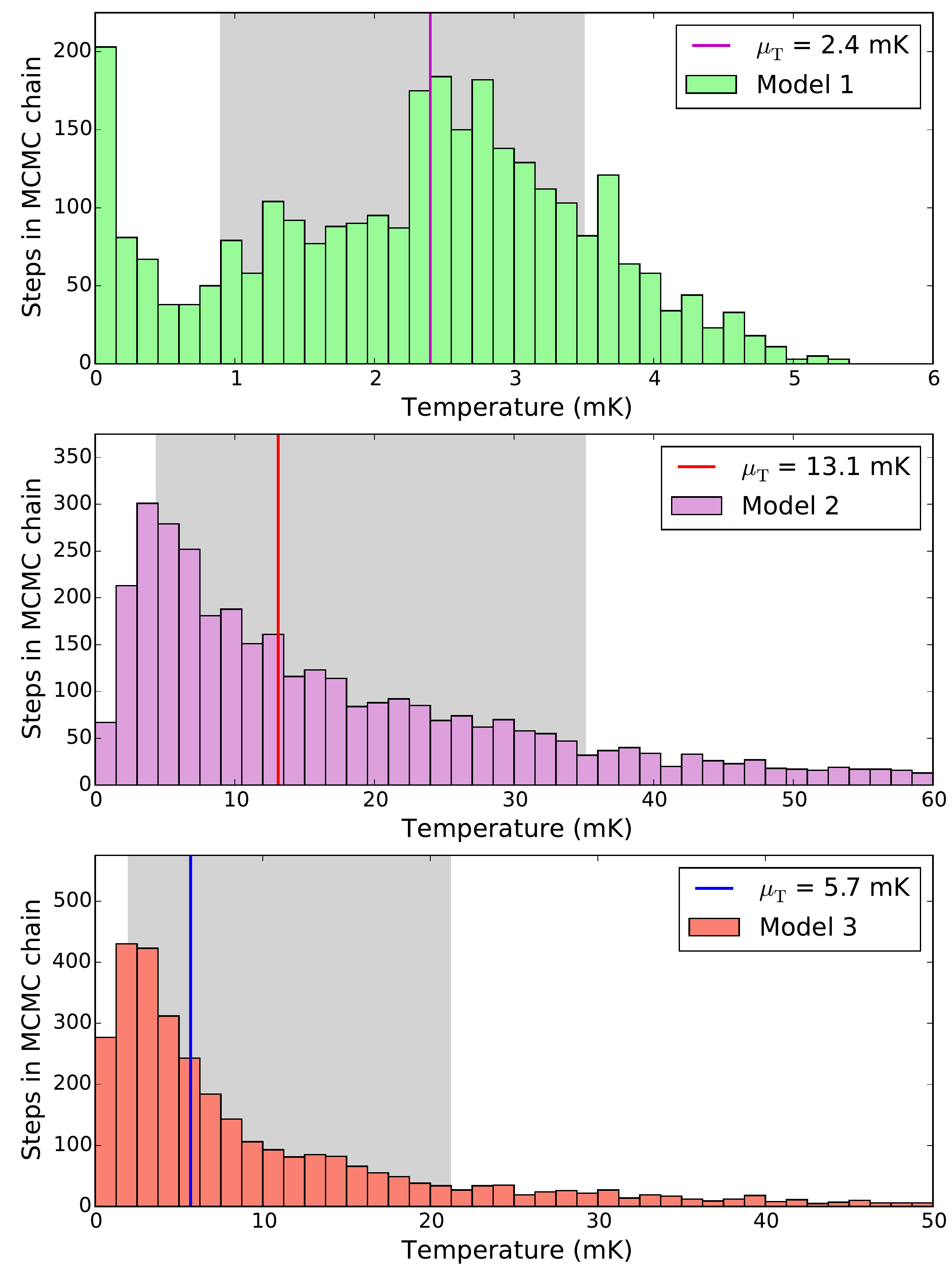}
\caption{Histograms of the contribution to the background temperature from MCMC fitting of the three source count models. Temperatures are for extended emission counts only, with the discrete source count being $T_{\rm dis}=63\,$mK. The top panel shows the background temperatures from fitting Model 1, the middle panel is the histogram from Model 2, and the bottom panel from Model 3. The solid vertical lines are the medians for each distribution, with the grey shaded regions showing the $68\,$per cent confidence regions.}
\label{fig:temphist}
\end{figure}

\begin{figure*}
\includegraphics[scale=0.5,natwidth=14in,natheight=10in]{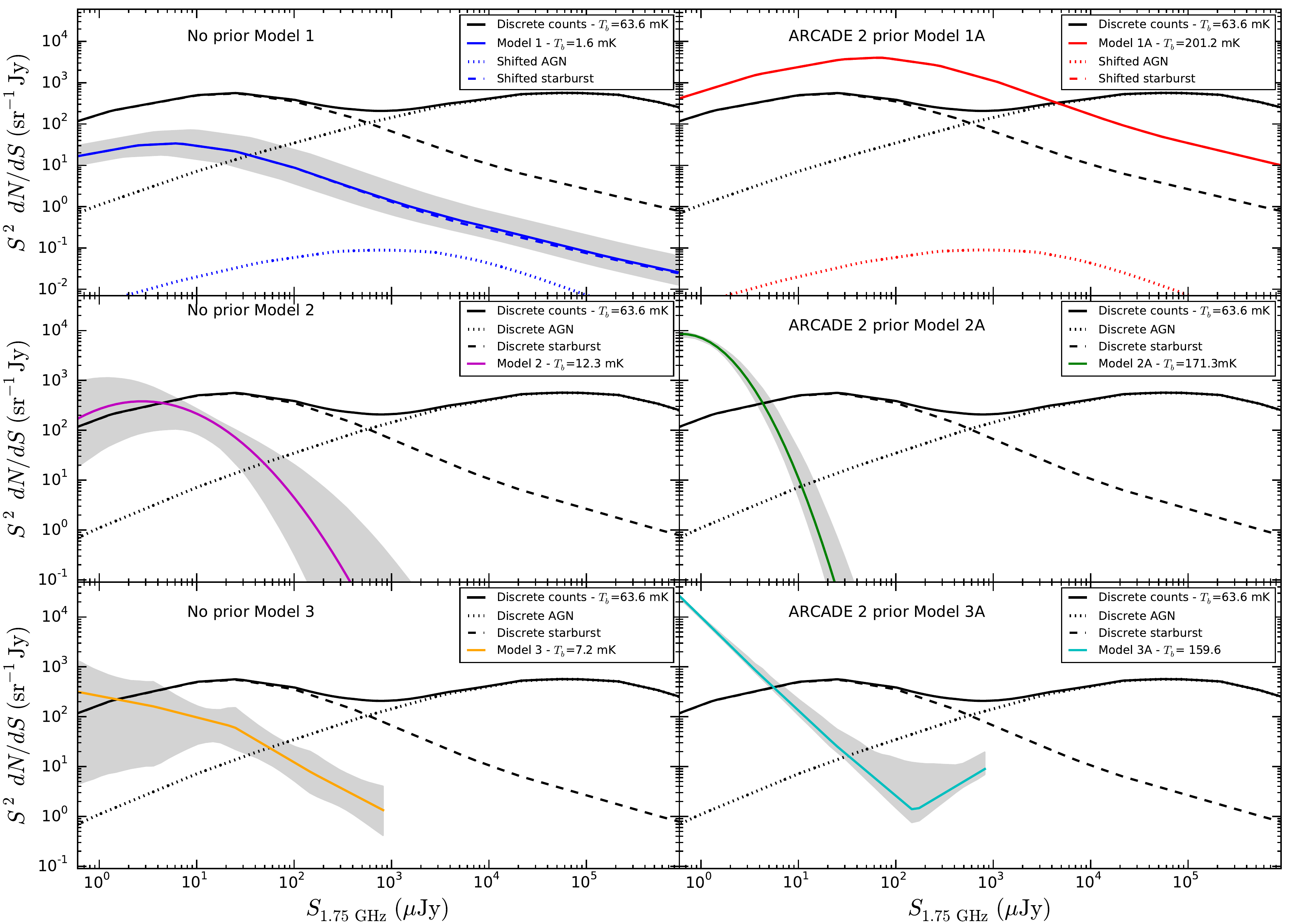}
\caption{$S^2$ normalized source counts at $1.75\,$GHz. The black lines are the same in all plots and are counts of discrete sources from the recent estimates by \citetalias{Vernstrom13} at $3\,$GHz, scaled to $1.75\,$GHz using $\alpha=-0.7$, while the coloured lines are the extended emission counts from model fitting. The count is broken into two populations, AGN and starbursts, based on evolutionary models, shown as the dotted and dashed lines, respectively, with the solid lines being the sum of all components. The top left panel shows Model 1 (blue lines), while the top right panel shows Model 1A (red lines). Note that the $S^2$ normalization makes it hard to see that the shifted populations have the same $dN/dS$ heights. The middle panels are Model 2 (left, purple lines) and Model 2A (right, green lines). The bottom panels are Model 3 (left, solid orange line) and Model 3A (light blue solid line). The shaded regions are the $68\,$per cent confidence intervals.}
\label{fig:counts_sh}
\end{figure*}

Each step in the MCMC chains is another source count model. For each model we used the MCMC results and calculated the background temperature distributions, using eq.~(\ref{eq:tint}), which are plotted in Fig.~\ref{fig:temphist}. The temperature distributions imply a mean temperature of $(10\pm 7)\,$mK. The resulting source count models are presented in Fig.~\ref{fig:counts_sh}, broken down by population and shown along with the discrete counts at $1.75\,$GHz.

 \begin{table}
 \centering
\caption{Best-fitting results for Model 1. The temperature, $T_{\rm ext}$, is the contribution to the background (using eq.~\ref{eq:tint}), for the extended source count contribution only. }
\begin{tabular}{lrr}
\hline
\hline
 & Model 1& Model 1A \\
 & (unconstrained) & (constrained)\\
 \hline
 $\log_{10}[C_1]$&$-1.91\pm0.11$&$-2.0\pm0.15$\\
 $\log_{10}[C_2]$&$-0.61\pm0.16$&$0.39\pm0.03$\\
$T_{\rm ext}$ { (mK)}& $1.65_{-0.75}^{+1.85}$ &$201.2\pm40$\\
$\sigma_{\rm c}$ {( $\mu$Jy beam$^{-1}$)}&$62.63$&$480.1$\\
$\sigma_{\rm c}$ {(mK)}&$2.78$&$21.36$\\
$\chi^2$ { ($N_{\rm dof}=42$)}&$109.6$&$45200$\\
 \hline
  \end{tabular}
 \label{tab:shifts}
 \end{table} 
 
  \begin{table}
 \centering
  \caption{Best-fitting parameter results for Model 2 and Model 2A. The temperature, $T_{\rm ext}$, is the contribution to the background (using eq.~\ref{eq:tint}), for the extended source count contribution only. }
  \begin{tabular}{lrr}
\hline 
\hline
Parameter & Model 2&Model 2A \\
 & (unconstrained) & (constrained)\\
\hline
$A$& $-0.79\pm0.29$&$-2.04\pm0.22$\\
$h$ & $ -5.55\pm0.40$&$-6.19\pm0.04$\\
$k$& $2.58\pm0.49$ &$3.93\pm0.05$ \\
\hline
$T_{\rm ext}$ {(mK)}  & $12.3_{-7.90}^{+22.8}$&$171.3_{-13.3}^{+16.2}$\\
$\sigma_{\rm c} \,${ ($\mu$Jy beam$^{-1}$) } &$62.81$ &$63.10$\\
$\sigma_{\rm c} \,${ (mK) } &$2.79$ &$2.81$\\
$\chi^2${ ($N_{\rm dof}=41$)}&$76.1$&$111.1$\\
\hline
\end{tabular}
\label{tab:p3pop}
\end{table}

  \begin{table}
 \centering
  \caption{Best-fitting parameter results for Model 3 and Model 3A.  The temperature, $T_{\rm ext}$, is the contribution to the background (using eq.~\ref{eq:tint}), for the extended source count contribution only. }
  \begin{tabular}{lrr}
\hline 
\hline
Parameter & Model 3&Model 3A\\
 & (unconstrained) & (constrained)\\
 \hline
$\log_{10}[\frac{S} {{\rm Jy}}]$ & $\log_{10}[\frac{dN/dS}{{\rm sr}^{-1} \, {\rm Jy}^{-1}}]$&$\log_{10}[\frac{dN/dS}{{\rm sr}^{-1} \, {\rm Jy}^{-1}}]$\\
\hline
$-6.25$ &$15.01\pm1.26$&$16.97\pm0.04$\\
$-5.43$ &$13.06\pm0.74$&$13.77\pm0.07$\\
$-4.62$ &$11.04\pm0.62$&$10.67\pm0.13$\\
$-3.81$ &$8.50\pm0.76$&$7.73\pm0.55$\\
$-3.00$ &$6.04\pm0.92$&$7.05\pm0.17$\\
\hline
$T_{\rm ext}$ {(mK)}  & $7.2_{-5.20}^{+14.0}$&$159.6_{-12.6}^{+9.50}$\\
$\sigma_{\rm c} \,${($\mu$Jy beam$^{-1}$) } &$62.73$ &$78.12$\\
$\sigma_{\rm c} \,${(mK) } &$2.79$ &$3.47$\\
$\chi^2$ {($N_{\rm dof}=39$)}&$75.3$&$251.6$\\
\hline
\end{tabular}
\label{tab:pmnode}
\end{table}
 
\subsection{Model uncertainties}
\label{sec:system}

We tried variations in the fitting method by first changing the fit statistic used ($\chi^2$ vs. $\log$ likelihood), which produced little change in the output; and second by trying different models. Instead of the parabola we tried a Gaussian in $S^2{\rm d}N/{\rm d}S$. The Gaussian model produces a peak in roughly the same spot as the parabola, though the parameters are not as well constrained. All models tried resulted in best-fit parameters that yielded background temperature estimates for the extended emission in the range of $(10 \pm 7) \, $mK.

We tested whether an incorrect estimate of the instrumental noise of $(52 \pm 5) \, \mu$Jy beam$^{-1}$ could affect the results by re-fitting the models while allowing the noise to vary between $40$ and $70 \, \mu$Jy beam$^{-1}$. This has little effect, except in Model 3, where the faintest node is degenerate with the noise. Thus a higher noise would decrease the amplitude of the faintest node. Nevertheless, we conclude that our noise estimate cannot be far enough off to explain the excess {\pd} width.

\subsection{ARCADE 2 fits}
\label{sec:arc2f}
We refitted the models to explore what source counts would be necessary to yield background temperatures in the range predicted by ARCADE 2, and to assess how well those source counts fit our data. It is clear from Fig.~\ref{fig:temps} that shifting the two populations with the ARCADE 2 prior (Model 1A) is strongly inconsistent with our data. With Model 2A or Model 3A it is possible to obtain source count temperatures in the ARCADE 2 range and find a reasonable fit to our data. Figure~\ref{fig:counts_sh} shows that in doing so, such a population would need to be extremely faint and numerous. The typical flux density of the peak of the parabola is three orders of magnitude below our instrumental and confusion noise limits. That region of the source count is nearly impossible to constrain with existing data. With Model 3A, the fitting routine makes the faintest node higher in amplitude, since changes to the counts that far below the noise result in very little change in the predicted {\pd} shape. 

The two models are also difficult to interpret in terms of physical objects. Since these extended, faint, numerous objects would completely overlap on the sky, modelling them as discrete objects fails. Future work will examine whether a faint diffuse cosmic web structure could produce this emission.

We conclude that there are no source count models, to a depth of $1\, \mu$Jy, that are consistent with both our data and the ARCADE 2 background temperature. Scaling our best-fitting discrete and extended source-count temperature ($70\,$mK) to the ARCADE 2 frequency of $3.3\,$GHz via a spectral index of $-0.7$ gives only $13\,$mK, compared with the nearly $55\,$mK result from ARCADE 2.

\citetalias{Vernstrom13} ruled out a new discrete population peaking brighter than $50\,$nJy. Combing that with our constraint on an extended population peaking above $1\, \mu$Jy indicates that the ARCADE 2 result  is highly unlikely to be due to extragalactic emission. Residual emission from subtraction of the Galactic component thus seems a more likely explanation for the  excess seen by the ARCADE 2 experiment. The contribution from extragalactic sources from ARCADE 2 depends on the model used for the subtraction of the Galactic component. \citet{Subrah13} showed that using a more realistic model of the Galaxy, as opposed to the plane parallel slab used by ARCADE 2 \citep{Kogut11}, yields no excess in the extragalactic component over that estimated from source counts. 

\section{Discussion}
\label{sec:dis}

When unconstrained by the ARCADE results, we find that Model 2 and Model 3 fit our data significantly better than Model 1 (an improved $\Delta \chi^2$ per degree of freedom of around $34$). Though the $\chi^2$s for Model 2 and Model 3 are still somewhat high, either model is a reasonable approximation to the data, at least when compared with the unsubtracted discrete model on its own, which has a $\chi^2=335$ for $44$ degrees of freedom. We now consider potential astrophysical sources of this emission. 

\subsection{Sources of diffuse emission}
\label{sec:exsources}

\begin{table*}
 \centering
  \caption{Luminosity and redshift estimates for Model 2. }
  \begin{tabular}{lllllllllll}
\hline
\hline
$S_{\rm peak}$ &$S_{-50\%}$ &$S_{+50\%}$&\multicolumn{2}{c}{$z_{\rm pk}=0.25$}&\multicolumn{2}{c}{$z_{\rm pk}=0.5$}&\multicolumn{2}{c}{$z_{\rm pk}=1$}&\multicolumn{2}{c}{$z_{\rm pk}=2$}\\
 & & & $\log_{10}[\frac{L_{1.4}}{{\rm W m}^{-2}}] $& $\Delta z$&$\log_{10}[\frac{L_{1.4}}{{\rm W m}^{-2}}] $&$\Delta z$&$\log_{10}[\frac{L_{1.4}}{{\rm W m}^{-2}}] $&$\Delta z$&$\log_{10}[\frac{L_{1.4}}{{\rm W m}^{-2}}] $&$\Delta z$\\
($\mu$Jy) & ($\mu$Jy)& ($\mu$Jy)& & & & & & &&\\
\hline
$2.8$&$0.7$&$11.0$&$19.7$& $0.38$&$20.5$& $0.79$&$21.4$& $1.65$&$22.2$& $3.44$\\

\hline
\end{tabular}
\label{tab:lums}
\end{table*}

Model 1, consisting of only shifts in the discrete counts, is considered here as an approximate representation of individual galaxy haloes. The best-fit results from this model should be considered only as upper limits for galactic haloes. This model on its own does not optimally fit the data. If there are other sources contributing to the measured {\pd}, the fitting process would push the shifts artificially high in an attempt to make the model as consistent with the data as possible. Also, this model falls apart when considering nearby galaxies that have been observed with single dish telescopes. According to the model, a $1\,$Jy nearby galaxy would have an extended halo of $250\,$mJy. This type of emission has not been seen around such sources. This implies that if all galaxies have some form of diffuse halo then the flux density of that halo cannot simply be a fraction of the discrete flux density. Models using the luminosity functions of the separate populations, where the halos may be a fraction of the point source luminosity, and or may have different evolution with redshift, would likely produce more consistent results. 

For any of the models, in order to be consistent with known constraints on the cosmic infrared background (CIB) the emission process(es) must not be linked directly to star formation rates. Moreover, as noted in Section~\ref{sec:srcsize}, this technique can only constrain sources that are roughly $2\,$arcmin or smaller. Thus, these models are valid only for objects in that size range. 

Additionally, since we assume the sources in these models are powered by synchrotron emission, we must also consider the associated X-ray emission and how that compares with the known cosmic X-ray background (CXB). The electrons that generate the synchrotron emission can inverse-Compton (IC) scatter off of CMB photons to generate X-ray emission, the brightness of which we can estimate as follows.

The synchrotron and IC power are related by,
\begin{equation}
\frac{L_{\rm IC}}{L_{\rm sync}}=\frac{U_{0}(1+z)^4}{U_{\rm B}}
\label{eq:sync_ic}
\end{equation}
\citep[see e.g.][]{Nath10}. Here $L_{\rm sync}$ and $L_{\rm IC}$ are the synchrotron and IC luminosities, $U_{\rm B}$ is the magnetic field energy density, and $U_{0}=4.2\times 10^{-14}\,$J m$^{-3}$ is the CMB energy density at $z=0$. Using the simplification that all of the sources are at the same redshift, we can calculate the IC flux density for a range of redshift and magnetic field values. We used redshifts of $0.01\le z \le 4$ and magnetic field values in the range of $0.01\, \mu$G $\le B \le 10\, \mu$G \citep[based on $B$ values for nearby clusters being around $1\, \mu$G,][]{Feretti12}.

When integrating the new source count (d$N$/d$S_{\rm IC}$) we obtain a range of values of the CXB for energies of $100$s of keV. \citet{Churazov07} presented observations of the CXB spectrum measured by INTEGRAL of $E^2dN/dE \le 15 \,$keV$^2$ s$^{-1}$ cm$^{-2}$ keV$^{-1}$ sr$^{-1}$ for $E\ge 100 \,$keV. In these units our models yield values of $10^{-13}$ to $5$ for $E^2dN/dE$, depending on the assumed redshift and magnetic field strength (larger values for lower $B$ and higher $z$). This shows that such models should not have a large impact on the X-ray background.

\subsection{Cluster emission}
\label{sec:clust}

If we are to assume that Model 2 (middle right panel of Fig.~\ref{fig:counts_sh}) represents astrophysical sources, we need to determine how they compare to known objects. Making some simple assumptions, we can calculate possible luminosities and redshifts. We chose several redshifts for the peak of the parabola and calculated the $K$-corrected $1.4\,$-GHz luminosity, assuming a spectral index of $\alpha=-0.7$. Then, assuming the objects all have the same intrinsic luminosity, we calculated the redshifts at which the counts have fallen to $50\,$per-cent of the peak. We did this for peak redshifts of $z=0.25,\,0.5,\, 1,$ and $2$; the results are listed in Table~\ref{tab:lums}.

It seems unlikely that this population could represent cluster emission from radio haloes or relics. The luminosity values for such objects, given in \citet{Feretti12}, are in the range of $23\le \log_{10}[L_{1.4}] \le 26$, several orders of magnitude larger than seen here. To date, we know of less than $100$ clusters that host giant or mini radio haloes \citep{Feretti12}. Extended radio emission in clusters has only been observed in high mass clusters ($\ge10^{14}{\rm M}_{\odot}$) at low redshift, and all with total $1.4\,$GHz flux densities in the 10s to 100s of mJy. 

There could of course be similar objects (relics, haloes, etc.) in smaller mass groups at higher redshifts that are contributing. \citet{Nurmi13}, using data from the SDSS survey, found that the majority of galaxies reside in intermediate mass groups, as opposed to large clusters. Stacking of subsamples of luminous X-ray clusters by \citet{Brown11} found a signal of diffuse radio emission below the radio upper limits on individual clusters. It is possible that there are clusters or groups that are more `radio quiet', below current detection thresholds \citep[e.g.][]{Brunetti11,Cassano12}. 

\citet{Zandanel14b} used a cosmological mock galaxy cluster catalogue, built from the MultiDark simulation \citep{Zandanel14a}, to investigate radio loud and radio quiet  halo populations. Their model, which assumes $10\,$per cent of clusters to have radio loud haloes, is a good fit \citep[see figure 5 of][]{Zandanel14b} to the observed radio cluster data from the NVSS survey \citep{Giovannini99}. The luminosity limit for the observed NVSS data is $\log_{10}[L_{1.4}\, ({\rm W \, Hz^{-1}}) ]\simeq 23.5$, while the simulation continues to a limit of $\log_{10}[L_{1.4}\, ({\rm W \, Hz^{-1}}) ]\simeq 20$.

\begin{figure}
\includegraphics[scale=0.37,natwidth=9in,natheight=11in]{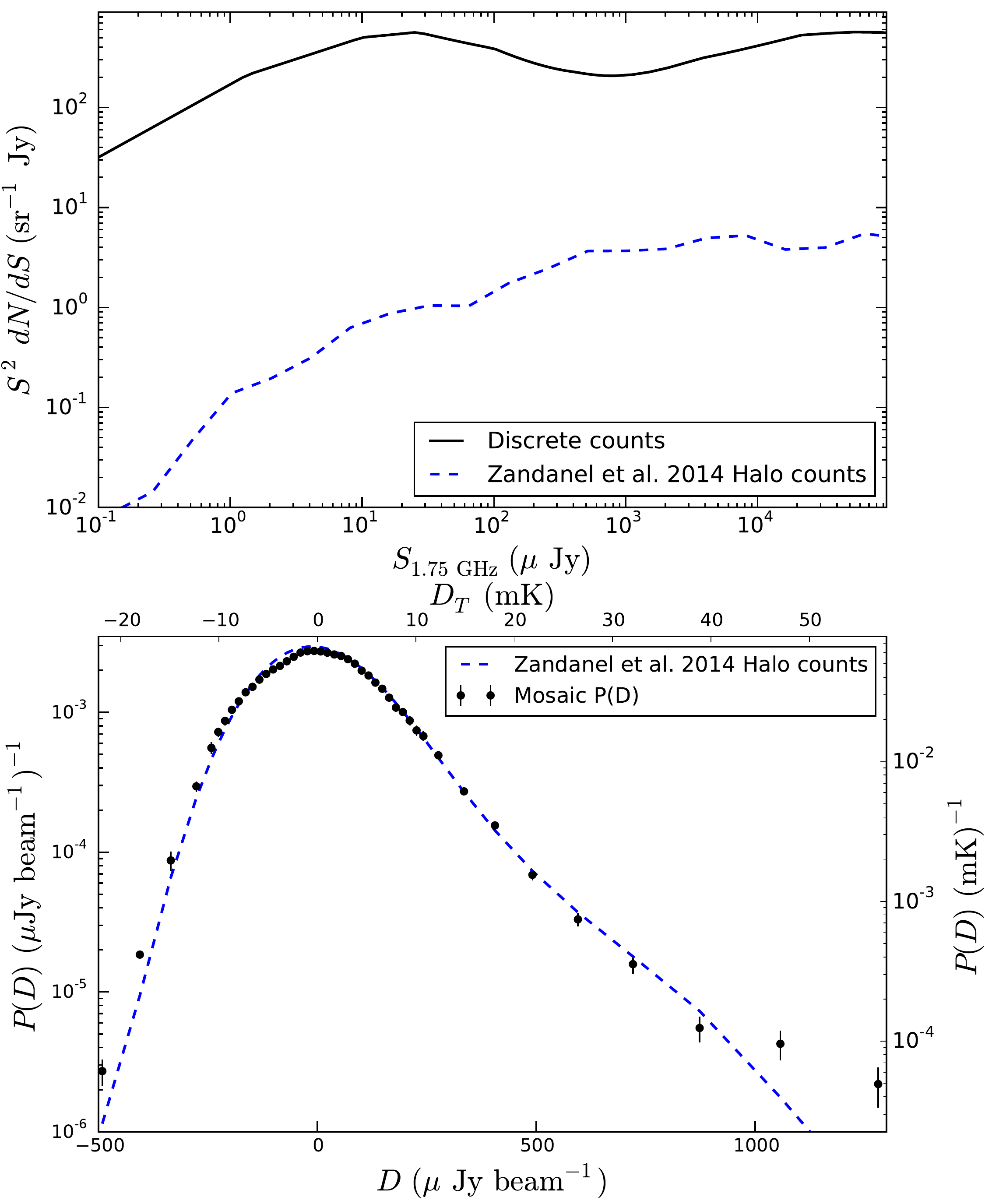}
\caption{Comparison of the radio cluster halo model from \citet{Zandanel14b} with current data. The top panel shows the $S^2$ normalized source count derived from taking the halo radio luminosity values at redshift snapshots  $z=0,\, 0.1,\, 0.5,$ and $1$, and converting to $1.75\,$GHz (blue dashed line), compared with the discrete radio source count (black solid line). The bottom panel shows the output {\pd} for the halo model plus the unsubtracted point source contribution, convolved with Gaussian noise of $52\, \mu$Jy beam$^{-1}$ (black points).  }
\label{fig:clft}
\end{figure}

It is instructive to compare these simulated haloes with our ATCA data. Using the online database to access the simulation \citep{Riebe13}\footnote{\url{http://www.cosmosim.org}}, we used the $1.4$-GHz halo simulation snapshots for $z=0,\, 0.1,\, 0.5,$ and $1$, scaling the luminosities for each redshift snapshot to give the flux density that would be observed at $1.75\,$GHz. We computed a source count from these data, combined it with the unsubtracted discrete emission model and Gaussian noise to obtain a predicted {\pd}. The source count and {\pd} are shown in Fig.~\ref{fig:clft}. The source count from this model only adds an additional $1.5\,$mK to the radio background temperature.

The fit to the image {\pd} is not unreasonable, with this model adding only a modest excess width to the distribution compared with the unsubtracted discrete model on its own. The source count would likely not decrease as significantly in the sub-mJy region if the simulation included data from redshifts higher than $z=1$, and this would likely improve the fit. The $\chi^2$ is high mainly due to this model having a slightly higher number of bright objects and thus over-predicting the tail of the distribution. However, some of these brighter haloes would be relatively nearby, hence larger on the sky and so potentially resolved out in our data (see Table~\ref{tab:zsizes}).

The halo model has similar count amplitude to our best-fit for Model 3 around $1\,$mJy. This halo model then begins to fall off, whereas the node model rises; this again could be due to the lack of high redshift objects. This type of halo model is therefore not necessarily inconsistent with our phenomenological model. Assuming the model from \citet{Zandanel14b} is a realistic extension of radio haloes to fainter luminosities, then it is possible for such haloes to exist given our data. However, more deep observations of clusters are necessary to test the accuracy of this model. 

One thing to keep in mind is that the model from \citet{Zandanel14b} deals with the issue of the origin of radio haloes, i.e. haloes being generated from re-acceleration or hadronic-induced emission. This model assumes a fraction of the observed radio emission is of hadronic origin. However, if the hadronic contribution is negligible, acting only at radio-quiet levels, the predicted counts would be dramatically lower at all masses.

\subsection{Dark matter constraints}
\label{sec:dmc}
It has been proposed that radio emission may originate from WIMP dark matter particles. Dark-matter particle annihilation in haloes releases energy as charged particles, which emit synchrotron radiation due to the magnetic field of the surrounding galaxy or galaxies. The predicted emission depends on the mass of the dark matter particle and halo mass or density profile, as well as the strength of the magnetic field. 

\begin{figure}
\includegraphics[scale=0.37,natwidth=9in,natheight=11in]{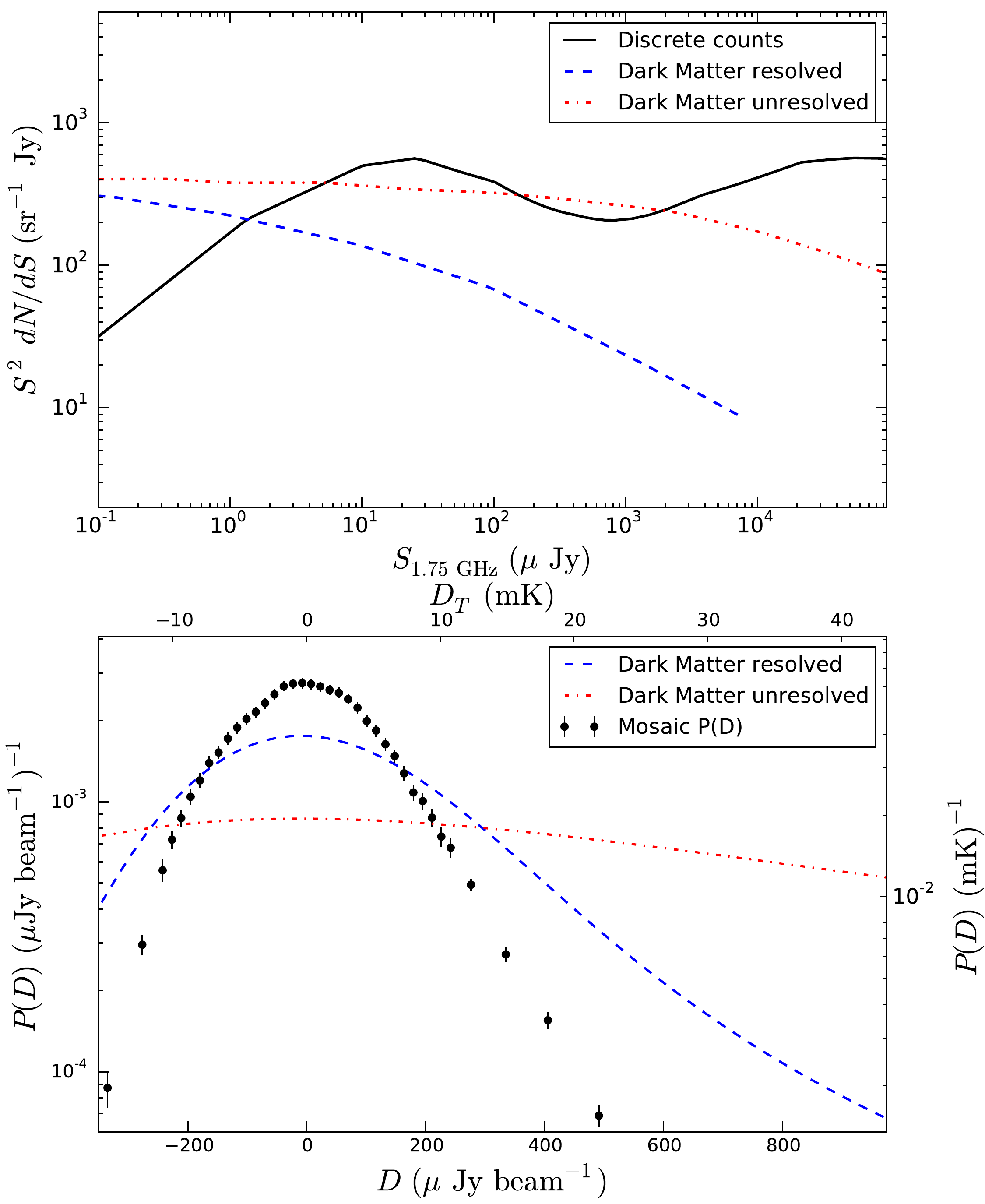}
\caption{Comparison of one particular dark matter model with current radio data. The top panel shows the two predicted source count models \citep[see figure 3 in][]{Fornengo11},shifted to $1.75\,$GHz, for a $10\,$GeV dark matter particle mass, assuming all the structures are resolved (blue dashed line) and unresolved (red dot-dashed line), together with the discrete radio source count (black solid line), with $S^2$ normalization. The bottom panel shows the output \pd s for the two models plus the unsubtracted point source contribution, convolved with Gaussian noise of $52\, \mu$Jy beam$^{-1}$ (black points).  }
\label{fig:dm3}
\end{figure}

\citet{Fornengo11} presented one dark matter model with two source-count predictions, the first assuming all the halo substructures are resolved and the second assuming all the substructures are unresolved. The predicted source counts, shifted to $1.75 \,$GHz, is shown in the top panel of Fig.~\ref{fig:dm3} along with the discrete radio source count. Their best-fit model has a dark matter mass of $10\,$GeV, assuming annihilation or decay into leptons. We computed the predicted {\pd} for both models, plus the unsubtracted discrete source contribution convolved with Gaussian noise of $52\, \mu$Jy beam$^{-1}$. The model \pd s are shown in the bottom panel of Fig.~\ref{fig:dm3} along with our radio image {\pd}. 

Clearly these particular models are not consistent with our current radio data.  Any other dark matter models would need reduced amplitude of the counts for flux densities greater than about $10\, \mu$Jy. Models with the dark matter count amplitude as high as or higher than that from known radio sources for these brighter flux densities would overproduce the emission seen and are therefore ruled out. Dark matter models consistent with our data and responsible for the ARCADE 2 emission would need to produce a large portion of the emission from the sub-$\mu$Jy region, a region not constrained by our data. However, the required number counts would render such predictions unrelated to galactic haloes.

\section{Integral counts}
\label{sec:intc}

\begin{figure*}
\includegraphics[scale=0.64,natwidth=11in,natheight=11in]{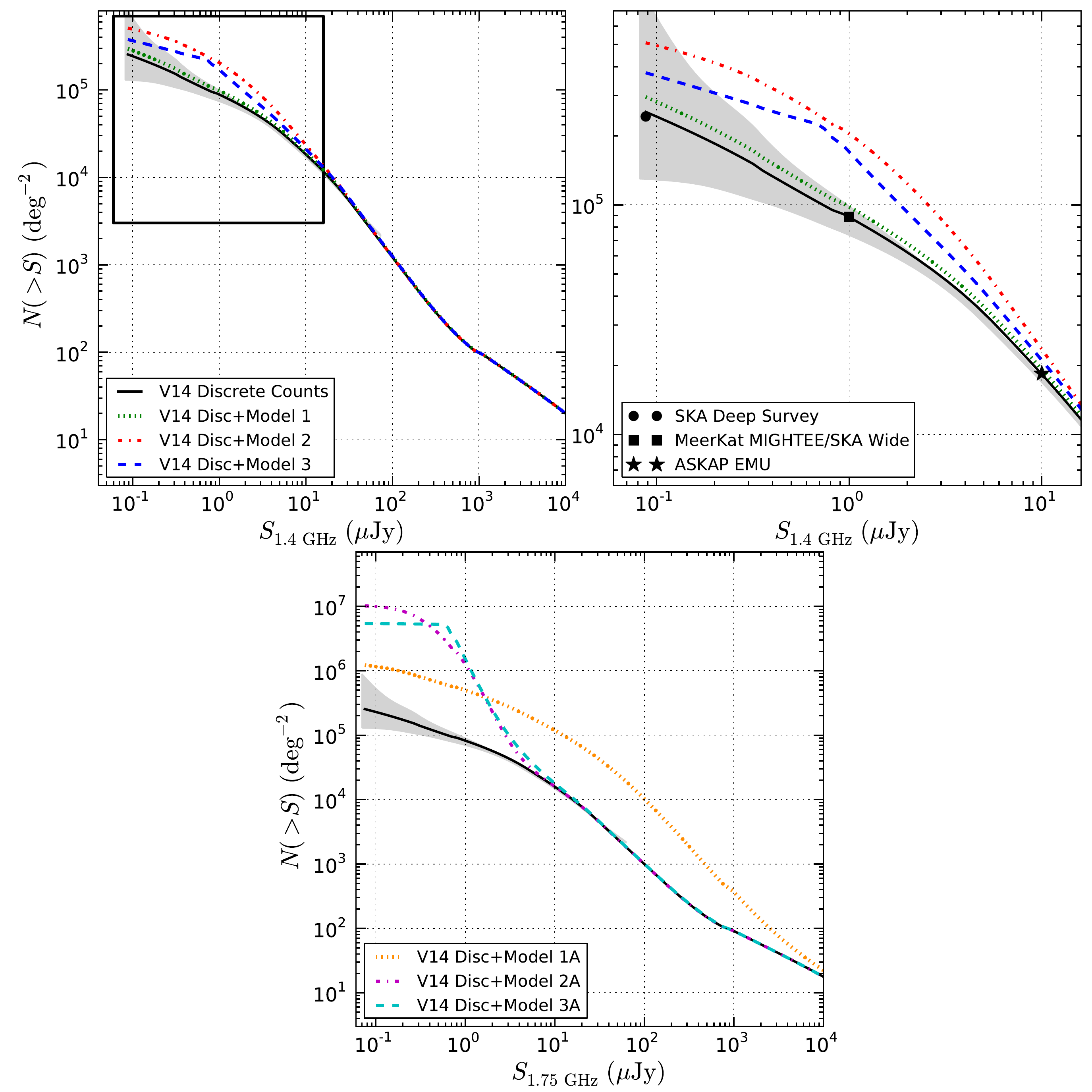}
\caption{Integrated source counts, or number of sources per square degree, at $1.4\,$GHz (top) and $1.75\,$GHz (bottom). The solid black lines are the discrete source count from \citetalias{Vernstrom13}, originally at $3\,$GHz, and scaled to 1.75 and $1.4\,$GHz using $\alpha=-0.7$. The green dotted lines are the discrete count plus Model 1. The red dot-dashed lines are the discrete count with the addition of Model 2. The blue dashed lines are the discrete source count with the addition of Model 3. The shaded grey areas represent $68\,$per cent confidence regions of the discrete count derived from \citetalias{Vernstrom13}. The upper right-hand panel shows a close up of the region marked by the solid rectangle in the upper left panel. The three points show the expected number of sources per square degree for the upcoming SKA and SKA Pathfinder surveys based on their expected depths (the circle is SKA, the square is MeerKAT MIGHTEE and the all sky SKA, and the star is the ASKAP EMU survey). The bottom panel ($1.75\,$GHz) shows Model 1A, Model 2A, and Model 3A as orange dotted, magenta dot-dashed, and light blue dashed lines, respectively (same models fit with the ARCADE 2 temperature prior).}
\label{fig:intcsd}
\end{figure*}

Now that we have closed the loophole of extended emission, we can revisit source count constraints in general. It is important for future deep survey designs to have an accurate estimate of the expected number of source detections. To estimate this we can derive the integral source counts $N(>S)$, or the total number of sources with flux density greater than $S$ per unit area. 
Deep and accurate estimates of $N(>S)$ can provide useful information for surveys at a range of frequencies, with proper scaling; in the synchrotron-dominated regime we should be able to extrapolate by a factor of a $\sim \pm 2$ in frequency. 

This is of particular relevance to the Square Kilometre Array (SKA), and its pathfinders, ASKAP and MeerKAT, as well as the new planned deep VLA survey. The VLA Sky Survey (VLASS) is aiming to map an area $10\,$deg$^2$ to a depth of $1.5 \, \mu$Jy at $1.4\,$GHz with a resolution of roughly $1\,$arcsec \citep{Jarvis14}. The Evolutionary Map of the Universe continuum survey \citep[EMU]{Norris11} planned for ASKAP will cover the entire sky south of Dec $+30^\circ$ with a resolution of $10\,$arcsec at $1.4\,$GHz, and will also be sensitive to diffuse emission with a sensitivity at $1\,$arcmin scale similar to that reached in this paper. The deep survey with MeerKAT \citep[MIGHTEE,][]{Jarvis11}, will reach an rms of $1\, \mu$Jy over $1000\,$deg$^2$ with arcsec resolution. In the following decade, the SKA will conduct an all-sky survey to an rms of $1\, \mu$Jy, and a smaller survey to an rms of $100\,$nJy. It would be helpful in planning to know what source densities are expected in these surveys.
 
We can obtain the integral source counts from
\begin{equation}
N(>S)=\int_S^{\infty} \frac{dN}{dS}dS.
\label{eq:intcte}
\end{equation}
We have derived the integrated source counts from the discrete model in \citetalias{Vernstrom13}, as well as that discrete model plus the best-fits from the extended emission models. These are shown in Fig.~\ref{fig:intcsd}, with values listed in Table~\ref{tab:intctab}. Also shown on the plot are the expected SKA and SKA Pathfinder survey limits. 

The SKA and Pathfinders should not be limited by any natural source confusion for discrete sources. The natural confusion limit is the confusion caused by the finite source sizes, as opposed to confusion caused by the telescope beam size. For discrete sources with an average source size of approximately $1\,$arcsec$^2$ for faint sources, the natural confusion limit would be less than $10\,$nJy. However,  extended objects of $2\,$arcmin diameter, for example, would begin to heavily overlap above $1000\,$sources per deg$^2$ which corresponds to a flux density at $1.4\,$GHz of approximately $100\,\mu$Jy.

To highlight some numbers (ignoring extended emission now) the discrete model predicts $1\times10^9$ sources over the whole sky brighter than $23\, \mu$Jy, and $10\,$sources per arcmin$^2$ brighter than $4.6\, \mu$Jy. At a limit of $1 \, \mu$Jy we estimate $88{,}500\,$sources per square degree at $1.4\,$GHz. For relatively modest extrapolations in flux density and frequency, the cumulative counts for $0.1\le S \le 5 \, \mu$Jy can be well described by 
\begin{equation}
N(>S)\simeq 84,800 \left ( \frac{S}{1{\rm \mu Jy}} \right)^{-0.48} \left ( \frac{\nu}{1.4{\rm GHz}} \right )^{-0.33} {\rm deg}^{-2},
\label{eq:nncnts1}
\end{equation}
and for $5< S \le 500 \, \mu$Jy 
\begin{equation}
N(>S)\simeq 296,700 \left ( \frac{S}{1{\rm \mu Jy}} \right)^{-1.20} \left ( \frac{\nu}{1.4{\rm GHz}} \right )^{-0.33} {\rm deg}^{-2}.
\label{eq:nncnts2}
\end{equation}

\begin{table}
\centering
\caption{Integrated source count values of the different models scaled to $1.4\,$GHz.}
  \begin{tabular}{lcccc}
\hline 
\hline
$\log_{10}[\frac{S}{\rm Jy}]$&Discrete& Dis+Mod1& Dis+Mod 2& Dis+Mod 3\\

& \scriptsize(No. deg$^{-2}$)& \scriptsize(No. deg$^{-2}$)& \scriptsize(No. deg$^{-2}$)& \scriptsize No. deg$^{-2}$)\\
\hline
$-7.0$&$2.4\times10^5$ &$2.8\times10^5$&$4.9\times10^5$&$3.6\times10^5$\\
$-6.5$&$1.5\times10^5$ &$1.7\times10^5$&$3.5\times10^5$&$2.7\times10^5$\\
$-6.0$&$8.8\times10^4$ &$9.8\times10^4$&$2.1\times10^5$&$1.7\times10^5$\\
$-5.5$&$4.7\times10^4$ &$5.1\times10^4$&$8.3\times10^4$&$6.3\times10^4$\\
$-5.0$&$1.8\times10^4$ &$1.9\times10^4$&$2.3\times10^4$&$2.1\times10^4$\\
$-4.5$&$5.4\times10^3$ &$5.6\times10^3$&$5.7\times10^3$&$5.8\times10^3$\\
$-4.0$&$1.2\times10^3$ &$1.2\times10^3$&$1.2\times10^3$&$1.2\times10^3$\\
$-3.5$&$2.9\times10^2$ &$2.9\times10^2$&$2.9\times10^2$&$2.9\times10^2$\\
\hline
\end{tabular}
\label{tab:intctab}
\end{table}

\section{Conclusions}
\label{sec:conclusion}
We used five antennas in the Australian Telescope Compact Array to observe seven pointings in the ELAIS-S1 field. Using these observations we constructed an image with a mean frequency of $1.75\,$GHz and a FWHM resolution of $150\,$arcsec$\,\times \, 60\,$arcsec. We performed subtraction of point source emission in the \textit{uv}-plane using models from the ATLAS survey, removing the discrete emission contribution above $S\simeq 150 \, \mu$Jy beam$^{-1}$. The image is confusion-limited with an rms of $(155\pm 5) \, \mu$Jy beam$^{-1}=(6.9\pm0.2)\,$mK and average instrumental noise $\sigma_{\rm n} =(52 \pm 5)\, \mu$Jy beam$^{-1}=(2.3\pm0.2)\,$mK.  A model of the unsubtracted point-source emission convolved with the mean instrumental noise yields an rms width estimate of $\sigma_{\rm c \otimes n}=(135\pm12) \, \mu$Jy beam$^{-1}=(6.0\pm0.5)\,$mK. This leaves an excess distribution of width and uncertainty (determined from bootstrap analysis) of $(76\pm23) \, \mu$Jy beam$^{-1}=(3.4\pm1.0)\,$mK, a difference significant at the $0.5\,$per cent level, unaccounted for by discrete sources. We take this excess width as an upper limit to the additional confusion provided by extended emission.

We used three kinds of source count models to examine this excess: that of extended emission from individual galactic haloes of active galactic nuclei and star-forming galaxies; a new population in the form of a parabola in the $S^2$-normalized source counts; and a model of connected power laws. These approaches, when combined with instrumental noise and the unsubtracted point source model, fit the {\pd} distribution of the data more or less equally well. The models resulted in a background temperature from extended emission of $T_{\rm extended}= (10\pm7) \,$mK, giving an estimate for the upper limit of the total radio background temperature at $1.75\,$GHz from extragalactic emission of $T_{\rm b}=(73 \pm 10) \,$mK. This rules out the possibility that sources of extended emission could provide the ARCADE 2 sky brightness temperature at this frequency of $265\,$mK, with $>5\sigma$ significance down to a level of $1\, \mu$Jy. 

Looking ahead to future deep surveys, we presented deep integral source counts at $1.4\,$GHz from both discrete and extended emission models. These can be easily scaled to estimate deep counts at nearby frequencies.

The models used represent upper limits on the extended emission, and are valid for sources with angular size of approximately $2\,$arcmin or less. Assuming the excess is truly from extended emission, rather than data artefacts, we discussed some possible sources for the extended emission. These include individual galaxy haloes from starburst or AGN galaxies, haloes from another population such as dwarf spheroidals (or something unknown), and possibly some contribution from clusters (or smaller mass groups) through emission structures such as radio relics and haloes. 

Modelling is required to see if any known objects could produce similar source count shapes, either by means of `normal' emission, i.e. synchrotron emission from magnetic fields, or from something more exotic such as via dark matter annihilation. We showed an example of dark matter models from \citet{Fornengo11} and found them to be inconsistent with our data. It is clear that the resulting source count for any WIMP dark matter model, for source sizes up to $2\,$arcmin, must lie below the source count of current radio galaxies, at least for flux densities greater than around $10\, \mu$Jy. 

If there is a large number of faint diffuse sources causing additional radio background temperatures, as suggested by ARCADE 2, it is unlikely that the emission can be seen by current-generation telescopes. For further constraints, assuming a steep spectral index, the natural way to search for such sources would be with a similar resolution survey at a much lower frequency, e.g. 325 or $610\,$MHz. Although we have focused here entirely on the 1-point statistics (i.e. \pd) a different approach would be to study the 2-point statistics of the radio sky. Measurements of the radio angular power spectrum could provide constraints on the smoothness of the radio background and statistically measure the clustering of the CRB over a range of angular scales.

\section*{Acknowledgments}
We acknowledge the support of the Natural Sciences and Engineering Research Council (NSERC) of Canada. The Australia Telescope Compact Array is part of the Australia Telescope National Facility which is funded by the Commonwealth of Australia for operation as a National Facility managed by CSIRO. The authors would like to thank Thomas Franzen and Julie Banfield for their hard work on the ATLAS data.

\bibliographystyle{mn2e}
\bibliography{ms.bbl}

\bsp

\label{lastpage}
\end{document}